\renewcommand\d{\delta}

\newcommand{\diracslash}[1]{#1\llap{/\kern2pt}}

\newcommand{\be}{\begin{equation}}
\newcommand{\ee}{\end{equation}}
\newcommand{\bea}{\begin{eqnarray}}
\newcommand{\eea}{\end{eqnarray}}
\newcommand{\ba}[1]{\begin{array}{#1}}
\newcommand{\ea}{\end{array}}

\documentclass[prd,aps,floats,nofootinbib,tightenlines,showpacs]{revtex4}
\usepackage{epsfig,graphicx,pstricks}
\usepackage{psfrag}
\usepackage{color}
\usepackage{amsmath}
\usepackage{amsfonts}
\usepackage{amssymb}
\usepackage{textcomp}
\usepackage{multirow}
\usepackage{subfigure}
\usepackage{dcolumn}% Align table columns on decimal point
\usepackage{bm}
\usepackage{ulem} %for strike through i.e newcommand cut

\begin{document}

\title {Estimating transport coefficients of interacting pion gas with K-matrix cross sections}
\author{Guruprasad Kadam }
\email{guruprasadkadam18@gmail.com}
\affiliation{Department of Physics, Shivaji University, Kolhapur, Maharashtra, India}
\author{Swapnali Pawar}
\email{swapna191286@gmail.com}
\affiliation{Department of Physics, The New College, Kolhapur, Maharashtra, India}
 \author{Hiranmaya Mishra}
\email{hm@prl.res.in}
\affiliation{Theory Division, Physical Research Laboratory,
 Navarangpura, Ahmedabad - 380 009, India}
\date{\today} 

\def\be{\begin{equation}}
\def\ee{\end{equation}}
\def\bearr{\begin{eqnarray}}
\def\eearr{\end{eqnarray}}
\def\zbf#1{{\bf {#1}}}
\def\bfm#1{\mbox{\boldmath $#1$}}
\def\hf{\frac{1}{2}}
\def\sl{\hspace{-0.15cm}/}
\def\omit#1{_{\!\rlap{$\scriptscriptstyle \backslash$}
{\scriptscriptstyle #1}}}
\def\vec#1{\mathchoice
        {\mbox{\boldmath $#1$}}
        {\mbox{\boldmath $#1$}}
        {\mbox{\boldmath $\scriptstyle #1$}}
        {\mbox{\boldmath $\scriptscriptstyle #1$}}
}

\begin{abstract}
 We estimate the transport coefficients, $viz.$, shear and bulk viscosities as well as thermal and electrical conductivities, of hot pionic matter using relativistic Boltzmann equation in relaxation time approximation. We use K-matrix parametrization of pion-pion cross sections to estimate the transport coefficients which incorporate multiple heavy resonances while simultaneously preserving the unitarity of S-matrix. We compare transport coefficients estimated using K-matrix parametrization with existing literature on pionic transport coefficients. We find that the K-matrix scheme estimations are in reasonable agreement with previous results. 
\end{abstract}


\maketitle

\section{Introduction}
 One of the long standing challenge in the theoretical and experimental nuclear physics is to establish the structure of phase diagram of strongly interacting matter. Relativistic Heavy Ion Collider (RHIC) is one of the early experiment  to explore the properties of strongly interacting matter at very high temperatures. The surprising discovery made by physicists in this experiment is the collective flow exhibited by the outgoing hadrons\cite{Adler:2003kt,Adams:2003xp,Adcox:2003nr,Adams:2004bi}. This flow has been observed both in single particle transverse momentum distribution distribution as well as asymmetric azimuthal distribution. The former flow is called radial flow while later is called elliptic flow. Ideal fluid dynamics calculations reproduces the measurements of both radial and elliptic flow up to transverse momenta, $p_{T}\sim 1.5 $ GeV$/$c\cite{Huovinen:2001cy}.  Despite this experimental evidence theoretical calculations based on Anti-de-Sitter/conformal field theory (AdS/CFT) duality suggest that fluid cannot have zero viscosity. In fact, for any fluid that can be found in nature, the ratio of shear viscosity to entropy density cannot be smaller than the lower limit, $\frac{1}{4\pi}$\cite{Kovtun:2004de, Buchel:2003tz,Kovtun:2005ev,Son:2007vk,Kapusta:2008ng,Iqbal:2008by,Springer:2008js}. The argument based on kinetic theory and uncertainty principle also suggest the lower bound on $\eta/s$\cite{Danielewicz:1984ww}. This motivated many theoretical investigations of this ratio rigorously from microscopic theory ~\cite{Gavin:1985ph,Prakash:1993bt,Dobado:2003wr,Itakura:2007mx,Chen:2007xe,Dobado:2009ek,Demir:2008tr, Puglisi:2014pda, Kadam:2014cua,Kadam:2014xka,Kadam:2015xsa,Deb:2016myz,Singha:2017jmq,Kadam:2018hdo, Lang:2012tt,Ghosh:2014qba,Wiranata:2012br,Wiranata:2012vv,FernandezFraile:2009mi}. Further, it is found that the evolution matter produced in HIC is successfully described using dissipative relativistic hydrodynamics ~\cite{Gale:2013da,Schenke:2011qd,Shen:2012vn,Kolb:2003dz,Teaney:2000cw,DelZanna:2013eua,Karpenko:2013wva,
Holopainen:2011hq,Jaiswal:2015mxa}, and transport simulations~\cite{Xu:2004mz,Bouras:2010hm,Bouras:2012mh,Fochler:2010wn,Wesp:2011yy,Uphoff:2012gb,Greif:2013bb}.  This implies that relatively good agreement between ideal fluid calculations and RHIC data suggest small value of viscosity and not zero and it needs to be taken into account for the accurate description of the matter produced in heavy-ion collision. Further, there are strong theoretical and experimental evidences which suggest that $\eta/s$ should have minimum at hadron to quark-gluon-plasma phase transition point\cite{ Csernai:2006zz,Dobado:2008vt,Dobado:2009ek}. Thus the theoretical estimations of shear viscosity coefficient is of phenomenological importance in the context of heavy-ion collision experiments.  
 
 The bulk viscosity governs the equilibration of the system subjected to compression or dilatational perturbations. Although its magnitude is very small compared to shear viscosity coefficient its importance  in heavy ion collisions has recently been realized and cannot be neglected. Bulk viscosity vanishes for conformally symmetric system according to Kubo formula. Matter created in relativistic heavy ion collision at very high center-of-mass energies and in the initial stages of its evolution is conformally symmetric. Thus it is expected that bulk viscosity vanishes or it is very small in magnitude for such matter. But recent studies indicate that near hadron-QGP phase transition point the trace anomaly $(\epsilon-3P)/T$, which is measure of conformal symmetry breaking, attains peak\cite{ Bazavov:2009zn,Cheng:2009zi,Borsanyi:2010cj}. Similar peak is expected in bulk viscosity coefficient near  hadron-QGP phase transition point. Indeed, few studies have shown such peaking behavior in $\zeta/s$\cite{Kharzeev:2007wb,Karsch:2007jc}. Further during the expansion of the fireball, when the temperature approaches the critical temperarture large $\zeta$ can give rise to different interesting
phenomena like cavitation when the pressure vanishes and hydrodynamic description breaks down \cite{Rajagopal:2009yw, Bhatt:2011kr}. The
effect of bulk viscosity on the particle spectra and flow coefficients have been investigated in Refs. \cite{Monnai:2009ad,Denicol:2009am,Dusling:2011fd}
while the interplay of shear and bulk viscosity coefficients have been studied in Refs. \cite{Song:2009rh,Noronha-Hostler:2013gga,Noronha-Hostler:2014dqa}. Recognizing the importance of the bulk viscosity in the heavy-ion collision phenomenology it has been estimated for both the hadronic and the partonic systems using various models ~\cite{Dobado:2011qu,Lu:2011df,Dobado:2001jf,Davesne:1995ms,FernandezFraile:2009mi,Chen:2007kx,NoronhaHostler:2008ju,Sasaki:2008um,FernandezFraile:2008vu,Dobado:2012zf,Ozvenchuk:2012kh,Gangopadhyaya:2016jrj,Chakraborty:2010fr,Sasaki:2008fg}.
 
  The thermal conduction involve relative flow between energy and baryon number. Matter created in RHIC and LHC (Large Hadron Collider) is mostly baryon free. Thus thermal conduction does not occur in such systems. Also  central collisions do not produce macroscopic  electric and magnetic field. So in such collisions electrical conductivity cannot play any role. But in the off-central collisions at small centre-of-mass energies  matter produced will be baryon rich and the spectators in such collisions would produce macroscopic electric and magnetic field which of the order of pion mass squared ($e{\bf B}=e{\bf E}\sim m_{\pi}^{2}$) within proper time $1-2$fm$/$c \cite{Tuchin:2013ie,Voronyuk:2011jd}. In such systems electrical as well thermal conductivities plays crucial role in the hydrodynamical evolution of matter created in heavy ion collisions.
  Several groups have studied the thermal and electrical electrical conductivity, including lattice QCD\cite{Aarts:2014nba}, the chiral perturbation theory~\cite{Fukushima:2008xe}, numerical  solution of the Boltzmann equation (BE)~\cite{Greif:2014oia, Puglisi:2014sha}, holography~\cite{Finazzo:2013efa}, transport models ~\cite{Cassing:2013iz,Steinert:2013fza}, Dyson Schwinger calculations~\cite{Qin:2013aaa}, a dynamical quasiparticle model~\cite{Marty:2013ita,Berrehrah:2014ysa}, quasiparticle model~\cite{Srivastava:2015via,Thakur:2017hfc}, Effective fugacity quasiparticle model~\cite{Mitra:2016zdw} and lattice gauge theory~\cite{Gupta:2004,Aarts:2007wj,Buividovich:2010tn,Ding:2010ga,Burnier:2012ts,Brandt:2012jc,Amato:2013naa}. All these studies aim to study conductivities in the QGP phase but some of these do extend below the transition
 temperature towards the hadron gas (HG). Recently, conductivities has been investigated  for a pion gas~\cite{Fernandez-Fraile2006} and for hot hadron gas~\cite{Greif:2016skc,Ghosh:2016yvt,Samanta:2017ohm}. It has also been studied in the framework of
 Polyakov-Nambu-Jona-Lasinio (PNJL) model~\cite{Saha:2017xjq} and in Polyakov-Quark-Meson (PQM) model~\cite{Singha:2017jmq}.
 
 In this work we estimate the transport coefficients, $viz.$, shear and bulk viscosities as well as thermal and electrical conductivities of hot pionic matter.  The expressions for all the transport coefficients can be obtained using relativistic Boltzmann equation in the relaxation time approximation. The method to obtain the transport coefficients from the Boltzmann equation is quite standard and can be found in many references\cite{chapmanbook,Cercignani,Groot}.  We estimate the relaxation time by calculating pion-pion elastic scattering cross section using  K-matrix parametrization similar to that of Ref.\cite{Wiranata:2013oaa}. The special feature of K-matrix formalism is that it incorporate multiple heavy resonances while simultaneously preserving the unitarity of S-matrix\cite{Chung:1995dx,Wigner:1946zz,Wigner:1947zz}. One can obtain the S-matrix once we know the K-matrix for the given scattering process. Finally the thermodynamical properties will be estimated using S-matrix formulation of statistical mechanics\cite{Dashen:1969ep}. This method has already been used to estimate the shear viscosity of hadron gas using Chapman-Enskog approximation method in Ref. \cite{Wiranata:2013oaa}. We use all the formulas required to calculate scattering cross section from Ref.\cite{Wiranata:2013oaa}.   But unlike Ref. \cite{Wiranata:2013oaa} we use averaged thermal relaxation time to estimate the transport coefficient which is rather a reasonable approximation\cite{Moroz:2013haa}.  We further study the effect of $\pi-K$, $\pi-\eta$ and $\pi-N$ interactions\cite{Oller:1998hw,Broniowski:2015oha} on the transport coefficients. These interactions are important in the context of pion distributions in heavy-ion collision experiment\cite{Huovinen:2016xxq} as well as in the study of strangeness fluctuations in strongly interacting matter\cite{Friman:2015zua}. It is to be noted that the K-matrix formalism we use lacks repulsive interactions. Recently, the K-matrix formalism has been extended to include attractive as well as repulsive interactions\cite{Dash:2018can,Dash:2018mep}. So it is possible to improve current work to include the repulsive interactions as well.

We organize the paper as follows. In Sect. II we give simple derivations of all the transport coefficients using relativistic Boltzmann equation. In Sect. III we briefly describe the thermodynamics of interacting hadron resonance gas model. In Sect. IV we discuss the results and finally in Sect. V summarize and conclude.   

\section{Transport coefficients in relaxation time approximation}
In this section we give brief derivation of all the transport coefficients starting from the Boltzmann equation. The derivation of shear viscosity, bulk viscosity and thermal conductivity is based on Ref.\cite{Gavin:1985ph} while the derivation of electrical conductivity is based on Ref.\cite{Puglisi:2014sha}. We rederive all the transport coefficients to make the paper self contained.
\subsection{Boltzmann equation}
The relativistic Boltzmann equation in presence of external force $F^{\mu}$ is\cite{Yagi,Cercignani,Israel:1979wp,Groot,Hosoya:1983xm}
\be
p^{\mu}\frac{\partial f_{p}(x,p)}{\partial x^{\mu}}+mF^{\mu}\frac{\partial f_{p}(x,p)}{\partial p^{\mu}}=\mathcal{C}[f_{p}(x,p)]
\label{Boltz_Eq}
\ee

The Boltzmann equation in the local rest frame and in the absence of external force can be written as
\be
\frac{\partial f_{p}(x,t)}{\partial t}+{v^{i}_{p}}\frac{\partial f_{p}(x,t)}{\partial x^{i}}=\frac{\mathcal{C}[f_{p}]}{p^{0}}
\label{Boltz_Eq1}
\ee

where $f_{p}$ is the distribution function, $v^{i}_{p}=\frac{p^i}{p^0}$ is the single particle velocity and  $\mathcal{C}[f_{p}]$ is called collision integral. Note that $f_{p}$ is not necessarily equilibrium distribution function (which we shall denote by $f_{p}^{eq}$). The collision term govern the rate of change of distribution function due to collisions and hence all the transport processes. In the relaxation time approximation (RTA) the collision term can be written as

\be
\mathcal{C}[f_{p}]\simeq -u^{\mu}p_{\mu}\frac{(f_{p}-f_{p}^{eq})}{\tau}
\label{rta}
\ee
where $u^{\mu}$ is the velocity four vector. In the local rest frame, $u^{\mu}=(1,\bf{0})$.
This approximation is based  on the assumption that the system is not very far from equilibrium. i.e
\be
f_{p}=f_{p}^{eq}+\d f_{p}; \: f_{p}^{eq}\gg \d f_{p}
\label{rta_assm}
\ee

Also the task of collisions is to bring the system back to its equilibrium state and the decay is exponential with relaxation time $\tau$. Substituting (\ref{rta}) and (\ref{rta_assm}) in Eq. (\ref{Boltz_Eq1})  we obtain Boltzmann equation in relaxation time approximation 

\be
\frac{\partial f_{p}^{eq}(x,t)}{\partial t}+{v^{i}_{p}}\frac{\partial f_{p}^{eq}(x,t)}{\partial x^{i}}\simeq-\frac{\d f_{p}}{\tau}
\label{Boltz_EqRTA}
\ee
In the absence of collisions the collision terms vanishes. The form of $\d f_{p}$ govern the dissipation (transport through collisions) in the system and we shall use above equation together with hydrodynamic description of fluid to derive transport coefficients.

\subsection{Shear viscosity}
Near local equilibrium a system of pions can be described by the local temperature $T$, the velocity $\bf{u}$, and the local chemical potential $\mu$. All these quantities are assumed to vary very slowly in space and time. In the local rest frame and near equilibrium the pions obey local distribution function
\be
f_{p}^{eq}=\frac{1}{e^{(E_{p}-p^{i}u_{i})/T}+1}
\label{bose_df}
\ee
apart from small correction term which we ignore.

The stress-energy tensor $T^{\mu\nu}$ for the pionic fluid can be written in terms of fluid velocity $u^i$ and its gradients. The spatial part of which is
\be
T^{ij}=(\epsilon+P) u^iu^j+P\d^{ij}-\eta\bigg(\frac{\partial u^i}{\partial x^j}+\frac{\partial u^j}{\partial x^i}\bigg)-(\zeta-\frac{2}{3}\eta)\frac{\partial u^k}{\partial x^{k}}\d^{ij}
\label{tmunu}
\ee

where $\epsilon$ is the energy density, $P$ is mechanical pressure and $\eta$ and $\zeta$ are shear and bulk viscosity coefficients respectively. The terms involving $\eta$ and $\zeta$ are  dissipative terms which govern the dissipation in system due to collisions. 
Dissipative part of energy momentum tensor can also be defined in terms of distribution function $f_{p}$. The spatial part stress-energy tensor is defined as
\be
T^{ij}=\int d\tilde{p}\:\frac{p^ip^j}{E_{p}}f_{p}
\ee
where $d\tilde{p}=g\frac{d^3p}{(2\pi)^3}$ and $g$ is the degeneracy factor. Substituting $f_p$ from Eq. (\ref{rta_assm}) we get

\be
T^{ij}=\int d\tilde{p}\:\frac{p^ip^j}{E_{p}}f_{p}^{eq}+\int d\tilde{p}\:\frac{p^ip^j}{E_{p}}\d f_{p}
\ee

First term can be identified with the ideal part while the second term can be identified with the dissipative part. i.e
\be
T_{id}^{ij}=\int d\tilde{p}\:\frac{p^ip^j}{E_{p}}f_{p}^{eq}
\ee
and
\be
T_{diss}^{ij}=\int d\tilde{p}\:\frac{p^ip^j}{E_{p}}\d f_{p}
\label{tmunu_dissi}
\ee
The coefficient of shear viscosity can be obtained from this dissipative part once we know $\d f_{p}$ which we have already found using Boltzmann equation in RTA (Eq. (\ref{Boltz_EqRTA})). To further simplify the calculation we assume streamline flow of fluid of the form $(u_{x}(y),0,0)$. For this flow one can  readily obtain coefficient of shear viscosity using equilibrium distribution function (\ref{bose_df}) as\cite{Gavin:1985ph}

\be
\eta(T)=\frac{1}{15T}\int d\tilde{p} \frac{p^4}{E^2}\tau(E) f_{p}(1+f_{p})
\label{eta}
\ee
Shear viscosity of gas with multiple species is just sum over contribution of individual species.
\be
\eta(T)=\frac{1}{15T}\sum_{a}\int d\tilde{p}_{a} \frac{p^4}{E_{a}^2}\tau_{a}(E_{a}) f_{p}(1+f_{p})
\label{etamulti}
\ee

\subsection{Bulk viscosity}
The bulk viscosity governs the equilibration of the system subjected to compression or dilatational perturbations. Taking traces of Eqs. (\ref{tmunu}) and (\ref{tmunu_dissi}) and equating we get
\be
(T_{diss})^{i}_{i}=-3\zeta\frac{\partial u^{i}}{\partial x^{i}}=-\int d\tilde{p}\frac{p^2}{E_{p}}\tau(E_{p})\bigg(\frac{\partial f_{p}^{eq}(x,t)}{\partial t}+{v^{i}_{p}}\frac{\partial f_{p}^{eq}(x,t)}{\partial x^{i}}\bigg)
\label{bulk1}
\ee

In the local rest frame we write conservation of energy and momentum respectively as
\be
\frac{\partial \epsilon}{\partial T}=-(\epsilon +P)\frac{\partial u^i}{\partial x^i}
\label{ener_cons}
\ee
\be
\frac{\partial u^i}{\partial t}=-\frac{1}{\epsilon+P}\frac{\partial P}{\partial x^{i}}
\ee

Taking derivatives of Eq. (\ref{bose_df}) we get
\be
\d f_{p}=\tau(E_{p})\: f_{p}(1+f_{p})\:\bigg(E_{p}\frac{(\epsilon+P)}{c_{v}T}\frac{\partial u^i}{\partial x^i}-\frac{p^k}{E_{p}}\frac{\partial}{\partial x^{k}}p^{l}u_{l}\bigg)
\label{df_bulk}
\ee
where $c_{v}=\frac{d \epsilon}{d T}$ is the specific heat. Substituting Eq. (\ref{df_bulk}) in Eq. (\ref{bulk1}) and using Eq. (\ref{ener_cons}) we arrive at expression for bulk viscosity\cite{Gavin:1985ph}
\be
\zeta(T)=\frac{1}{9T}\int d\tilde{p}\:\tau(E_{p})\bigg(\frac{\bf{p}^{2}}{E_{p}^{2}}-3v_{n}^{2}\bigg)^{2}f_{p}^{eq}(1+f_{p}^{eq})
\ee
where $v_{n}^{2}=\frac{dP}{d\epsilon}$ is the speed of sound at constant number density. For the multicomponent hadron gas the bulk viscosity is just a sum over the contribution from individual species as

\be
\zeta(T)=\sum_{a}\frac{1}{9T}\int d\tilde{p}_{a}\:\tau_{a}(E_{a})\bigg(\frac{\bf{p}^{2}}{E_{a}^{2}}-3v_{n}^{2}\bigg)^{2}f_{a}(E_{a}/T)
\label{zeta}
\ee
We note that bulk viscosity is always positive quantity as the integrand is either zero or positive definite.
\subsection{Thermal Conductivity}
For the pionic matter baryon number is zero. So the transport processes due to thermal conduction, which is related to flow of energy density and baryon number, would not occur. Further the pion interaction does not conserve the pion number as well since the number changing processes are allowed during pion-pion scattering processes as governed by effective Lagrangian. Nonetheless, at low temperature these number changing processes are strongly suppressed so that pion number is approximately conserve\cite{juanthesis}. Thus the transport process due to thermal conduction owing to energy flow and pion number can be defined.  

The thermal conduction occurs due the relative flow of energy with respect to pionic enthalpy ($w=\epsilon+P$). The heat flux is defined as
\be
q^{i}=T^{0i}-\frac{w}{n}j^{i}
\label{heat_flx1}
\ee
where $n$ is the pion number density. The energy flux ($T^{0i}$) and three current $j^i$ are defined respectively as
\be
T^{0i}=\int d\tilde{p}\:p^{i}f_{p}
\label{en_flx}
\ee
\be
j^{i}=\int d\tilde{p}\:\frac{p^{i}}{E_{p}}f_{p}
\label{current}
\ee
In the static situation ($u^{i}=0$) the heat flux ($\bf{q}$) is proportion to temperature gradient ($\nabla{T}$) and the proportionality constant is just thermal conductivity coefficient. In the component form 
\be
q^{i}=-\lambda\frac{\partial T}{\partial x^{i}}
\label{heat_flx2}
\ee
The negative sign emphasize the the fact the heat conduction occurs from region of higher to lower temperature according to second law of thermodynamics. Since the transport occurs due to $\d f_{p}$ part of the distribution function, comparing Eqs. (\ref{heat_flx1}) and (\ref{heat_flx2}) we get

\be
\lambda\frac{\partial T}{\partial x^{i}}=-\int d\tilde{p}\frac{p^i}{E_{p}}\bigg(E_{p}-\frac{w}{n}\bigg)\d f_{p}
\ee

Again, using (\ref{Boltz_EqRTA}) we get

\be
\lambda\frac{\partial T}{\partial x^{i}}=\int d\tilde{p}\frac{p^i}{E_{p}}\bigg(E_{p}-\frac{w}{n}\bigg)\tau(E)\frac{p^{j}}{E_{p}}\frac{\partial f_{p}}{\partial x^{j}}
\label{lamdfbidt}
\ee

Taking the spatial derivative of Bose distribution function (\ref{bose_df}) we get
\be
\frac{\partial f_{p}}{\partial x^{j}}=\frac{f_{p}(1+f_{p})}{T}\bigg(\frac{d\mu}{dT}+\frac{E_{p}-\mu}{T}\bigg)\frac{\partial T}{\partial x^{j}}
\label{dfbidx}
\ee
Substituting Eq. (\ref{dfbidx}) in to Eq. (\ref{lamdfbidt}) we get

\be
\lambda\frac{\partial T}{\partial x^{i}}=\int d\tilde{p}\frac{p^ip^j}{E_{p}^2}\bigg(E_{p}-\frac{w}{n}\bigg)^2\tau(E_{p})f_{p}(1+f_{p})\frac{\partial T}{\partial x^{j}}
\ee

 where we have used Gibbs-Duhem relation, $dP=w\frac{dT}{T}+nTd(\frac{\mu}{T})=0$.     Above equation can be further simplified using integral identities involving vector functions\cite{chapmanbook}

\be
\lambda\frac{\partial T}{\partial x^{i}}=\int d\tilde{p}\frac{p^2}{E_{p}^2}\bigg(E_{p}-\frac{w}{n}\bigg)^2\tau(E_{p})f_{p}(1+f_{p})\frac{\partial T}{\partial x^{i}}
\ee
comparing the coefficient of $\nabla_{i}{T}$ on both sides we arrive at the expression for thermal conductivity coefficient\cite{Gavin:1985ph} 
\be
\lambda=\int d\tilde{p}\frac{p^2}{E_{p}^2}\bigg(E_{p}-\frac{w}{n}\bigg)^2\tau(E_{p})f_{p}(1+f_{p})
\ee

Thermal conductivity of multicomponent hadron gas is just a sum over contribution from individual hadronic species.
\be
\lambda(T)=\sum_{a}\frac{1}{3T^2}\int d\tilde{p}_{a} \frac{p^2}{E_{a}^{2}}\tau_{a}(E_{a})(E_{a}-h)^2 f_{a}(E_{a}/T)
\label{lambda}
\ee
where $h=\frac{w}{n}$ is the heat function.

\subsection{Electrical Conductivity}

Boltzmann equation in presence of external electromagnetic field can be written as

\be
p^{\mu}\frac{\partial f_{p}(x,p)}{\partial x^{\mu}}+qF^{\mu\nu}\frac{\partial f_{p}(x,p)}{\partial p^{\mu}}=\mathcal{C}[f_{p}(x,p)]
\label{Boltz_Eq2}
\ee

where $F^{\mu\nu}$ is the electromagnetic field tensor. The DC electrical conductivity is the response of the system to external static electric field $(\bf{E})$. So considering only electric field components of $F^{\mu\nu}$ i.e $F^{0i}=-\bf{E}$ and $F^{i0}=\bf{E}$ we obtain 

\be
\frac{2q}{T}\:{\bf{E.p}}\:f_{p}^{eq}(1+f_{p}^{eq})=\frac{E_{p}\d f_{p}}{\tau}
\label{ele_rta}
\ee

where we have used RTA given by Eq. (\ref{rta}) of the collision term.

The coefficient of electrical conductivity is defined as
\be
\Delta j_{el}^{i}=\sigma_{el}E^{i}
\ee
where electric current density $j_{el}^{i}$ is defined as
\be
j_{el}^{i}=q\int d\tilde{p}\frac{p^{i}}{E_{p}}f_{p}
\ee
Thus

\bea
\sigma_{el}E^{i}&=&q\int d\tilde{p}\frac{p^{i}}{E_{p}}\d f_{p}\nonumber\\
&=&\frac{2q^2}{T}\int d\tilde{p}\frac{p^{i}p^{j}}{E^{2}_{p}}\:\tau(E_{p})\:f_{p}^{eq}(1+f_{p}^{eq})E^{j}\nonumber\\
&=&\frac{2q^2}{3T}\int d\tilde{p}\frac{p^{2}}{E^{2}_{p}}\:\tau(E_{p})\:f_{p}^{eq}(1+f_{p}^{eq})E^{i}
\eea
where in the last line we have used again the integral identity involving vector functions\cite{chapmanbook}.
Thus the coefficient of electrical conductivity is\cite{Puglisi:2014sha}
\be
\sigma_{el}(T)=\frac{q^2}{3T}\int d\tilde{p}\frac{p^{2}}{E^{2}_{p}}\:\tau(E_{p})\:f_{p}^{eq}(1+f_{p}^{eq})
\ee

For multicomponent hadron gas
\be
\sigma_{el}(T)=\sum_{a}\frac{q^2}{3T}\int d\tilde{p}_{a} \frac{p^2}{E_{a}^{2}}\tau_{a}(E_{a})f_{p}^{eq}(1+f_{p}^{eq})
\ee

 The relaxation time, $\tau_{a}(E_{a})$ is  in general energy dependent.  In this work we limit the estimations of transport coefficients within thermal averaged relaxation times defines as
\be
\tilde\tau_{a}^{-1}=\sum_{b}n_{b}\langle \sigma_{ab}v_{ab}\rangle
\label{therm_av}
\ee
where $n_b$ is the number density,  $\sigma_{ab}$ is the total cross section for the elastic scattering $ab\longrightarrow cd$ and $v_{ab}$ is the relative velocity. The angled bracket represents the thermal average. In the Boltzmann approximation the thermal average $\langle \sigma_{ab}v_{ab}\rangle$ can be calculated in the CM frame\cite{Cannoni:2013bza,Gondolo:1990dk}
\be
\langle \sigma_{ab}v_{ab}\rangle=\frac{1}{8Tm_{a}^{2}m_{b}^{2}K_{2}(m_a/T)K_{2}(m_{b}/T)}\int_{m_{a}+m_{b}}^{\infty}ds\frac{[s-(m_a-m_b)^2]}{\sqrt{s}}[s-(m_a+m_b)^2]K_{2}(\frac{\sqrt{s}}{T})\sigma_{T}(\sqrt{s})
\ee
So once we know the cross section for a given process the transport coefficients can be estimated through relaxation time given by Eq. (\ref{therm_av}).

 Also within the temperature range of interest ($100-150$MeV) Boltzmann approximation is rather good approximation whence we ignore Bose enhancement factors, i.e $f_{p}(1+f_{p})\simeq f_{p}\approx e^{-E_{p}/T}$.

\section{Thermodynamics of interacting hadron gas}

In this section we recapitulate the the thermodynamics of interacting hadron gas given in Ref.\cite{Welke:1990za,Venugopalan:1992hy}.  In the S-matrix formulation of statistical mechanics introduced  by Dashen, Ma and Bernstein\cite{Dashen:1969ep} the thermodynamical quantities can be computed if the S-matrix of the system is known. In this formulation the partition function can be written as 
\be
\text{ln}Z=\text{ln}Z_{0}+\sum_{i,j}z^{i}_{1}z^{j}_{2}(i,j)
\label{virial}
\ee
where $z=\text{exp}(\beta\mu)$ is the fugacity. In Eq.(\ref{virial}) the first term represent non-interacting part while the second term includes the interactions and the virial coefficients $b(i,j)$ which carries all the information about the interactions are written as\cite{Welke:1990za,Venugopalan:1992hy}
\be
b(i,j)=\frac{V}{4\pi i}\int\frac{d^3p}{(2\pi)^{3}}\int d\sqrt{s}\:\text{exp}[-\beta(p^2+s)^{1/2}]Tr_{i,j}\bigg[\mathcal{A}\mathcal{S}^{-1}(\sqrt{s})\frac{\vec{\partial}}{\partial \sqrt{s}}\mathcal{S}(\sqrt{s})\bigg]_{c}
\ee
where $\beta,V,p,\sqrt{s}$ represents inverse temperature, volume, total centre-of-mass momentum and centre-of-mass energy respectively. $i,j$ corresponds to a channel in $S$ matrix containing $i+j$ particle in the initial stage. $\mathcal{A}$ denotes the symmetrization (anti-symmetrization) operator for bosons (fermions). The trace is taken over all combinations of particle number and over all connected graphs.

At low temperatures ($T\leq m_{\pi}$) it is legitimate to assume that the hadrons interact mainly through elastic scattering. The S-matrix in this case is\cite{Venugopalan:1992hy}
\be
\mathcal{S}(\sqrt{s})=\sum_{I,l}(2l+1)(2I+1)e^{2\d^{I}_{l}}
\ee
where $I$ and $l$ corresponds to isospin channel and the angular momentum and $\d_{l}^{I}$ is the phase shift for $l^{th}$ partial wave. The second virial coefficient can be simplifies to get
\be
b_{2}=\frac{T}{2\pi}\int_{M}^{\infty}d\sqrt{s}\:s\:K_{2}(\sqrt{s}/T)\sum_{l,I}(2l+1)(2I+1)\frac{\partial\d^{I}_{l}(\sqrt{s})}{\partial\sqrt{s}}
\label{b2}
\ee
where $K_{n}$ is the modified Bessel's function of second kind and $M$ is the invariant mass of the interacting pair of particles at threshold. Note that in the summation for given $l$ sum over $I$ is restricted to the values consistent with the statistics. In the limit $\sqrt{s}\longrightarrow M$, $\delta_{l}^{I}\longrightarrow 0$ and Eq.(\ref{b2}) reduces to
\be
b_{2}=\frac{1}{2\pi^3}\int_{M}^{\infty}d\sqrt{s}\: s K_{1}({\sqrt{s}/T})\sum_{l,I}^{'}g_{l}^{I}\delta^{I}_{l}(\sqrt{s})
\ee
Thus the thermodynamical state variables of an interacting hadron gas is the sum of non-interacting part and the interacting part. Interacting part of the number density, energy density, pressure and the entropy density are respectively given by\cite{Welke:1990za,Venugopalan:1992hy}

\be
n_{int}=\frac{1}{\pi^3}\int_{M}^{\infty}d\sqrt{s}\: s K_{1}({\sqrt{s}/T})\sum_{l,I}^{'}g_{l}^{I}\delta^{I}_{l}(\sqrt{s})
\ee

\be
\epsilon_{int}=\frac{1}{4\pi^3}\int_{M}^{\infty}d\sqrt{s}\: s^{3/2} [K_{2}({\sqrt{s}/T})+K_{0}({\sqrt{s}/T})]\sum_{l,I}^{'}g_{l}^{I}\delta^{I}_{l}(\sqrt{s})
\ee

\be
P_{int}=\frac{T}{2\pi^3}\int_{M}^{\infty}d\sqrt{s}\: s K_{1}({\sqrt{s}/T})\sum_{l,I}^{'}g_{l}^{I}\delta^{I}_{l}(\sqrt{s})
\ee

\be
S_{int}=\frac{T}{2\pi^3}\int_{M}^{\infty}d\sqrt{s}\: s^{3/2} K_{2}({\sqrt{s}/T})\sum_{l,I}^{'}g_{l}^{I}\delta^{I}_{l}(\sqrt{s})
\ee

The phase shifts $\d^{I}_{l}$ carries all the information about the interaction and the knowledge of which determines the thermodynamics of hadron gas.

\section{Results and discussion}
Microscopically the magnitude of  transport coefficients is determined by the interaction strength between constituents of a system. This information is technically carried by scattering cross section $\sigma$ which determines the viscosity coefficients through the magnitude of relaxation time $\tau$. In case of hadronic matter, where the constituents are baryons and mesons, first principle calculations of hadronic cross sections do not exists but for the lightest hadrons. Thus, it is often useful to parametrize the cross section based on empirical information like phase shifts and decay widths. In this work we use two different cross section parametrization schemes, $viz.$, phase shift parametrization and K-matrix parametrization details of which are given in the appendix. For $\pi\pi\longrightarrow\pi\pi$  elastic scattering we consider all the resonance decay channels with appreciable branching ratios (see Table \ref{table}).  The particle data is taken from\cite{Beringer:1900zz}.

\begin{figure}[h!]
\vspace{-0.4cm}
\begin{center}
\begin{tabular}{c c}
 \includegraphics[width=8cm,height=8cm]{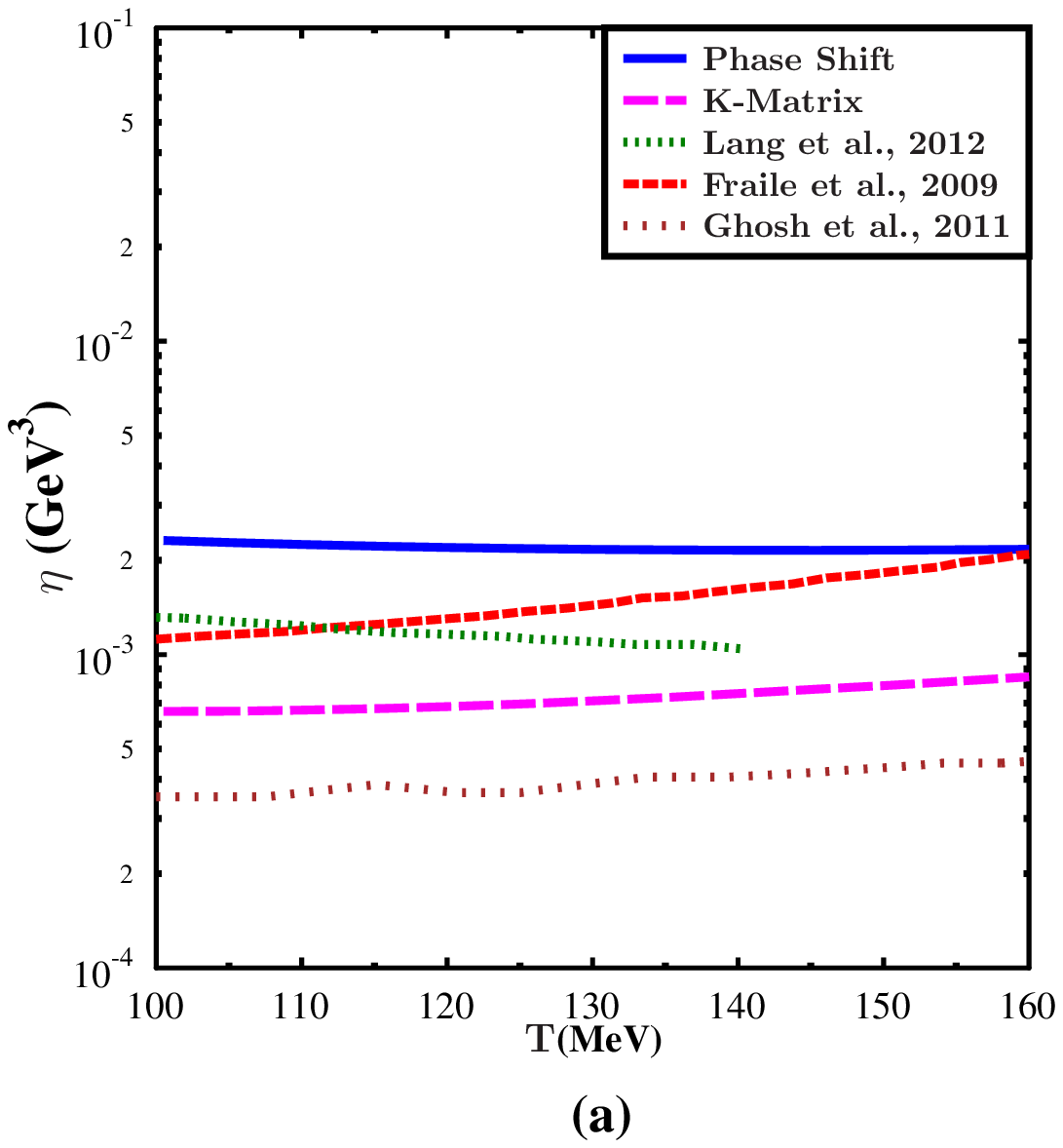}&
  \includegraphics[width=8cm,height=8cm]{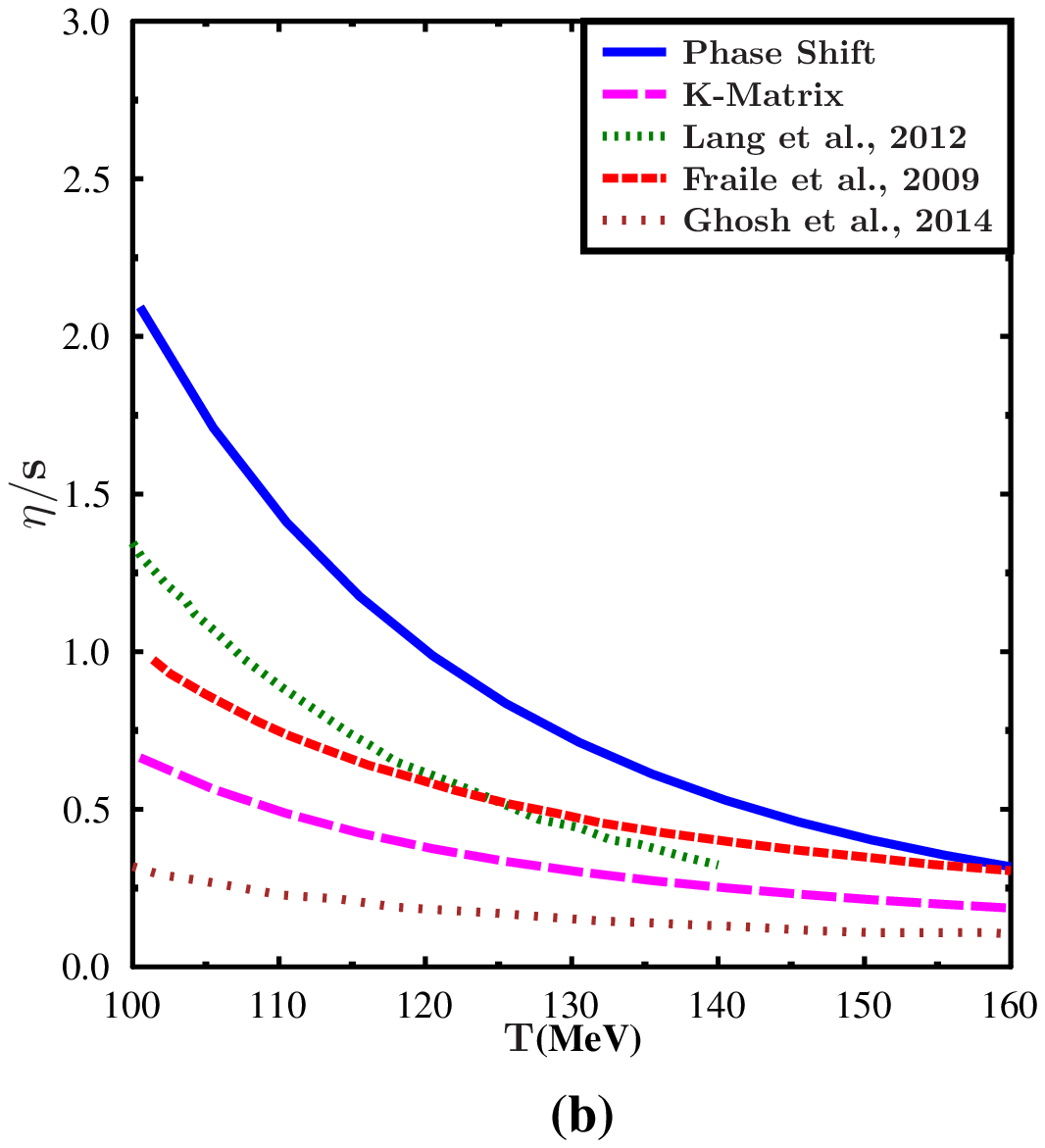}\\
  \end{tabular}
  \caption{Left panel shows shear viscosity estimate for pion gas  using phase shift and K-matrix parametrization of cross section used in this work  compared to previous results\cite{ Lang:2012tt,FernandezFraile:2009mi,Ghosh:2014qba}. Right panel shows ratio $\eta/s$ for pion gas compared to other works.} 
\label{shear_pion}
  \end{center}
 \end{figure}

 Fig. [\ref{shear_pion}(a)] shows shear viscosity as a function of temperature estimated using phase shift (solid blue) and K-matrix (dashed magenta) cross section parametrization. We have also compared our results  with other model estimations\cite{ Lang:2012tt,FernandezFraile:2009mi,Ghosh:2014qba}.  The shear viscosity coefficient itself is very small as far as order of magnitude is concerned. We note that estimations based on K-matrix parametrization are smaller  compared to the results of chiral perturbation theory\cite{Lang:2012tt,FernandezFraile:2009mi} while the estimations based on phase shift parametrization are larger. Fig. [\ref{shear_pion}(b)] shows the dimensionless ratio $\eta/s$ estimated within K-matrix formalism and various other models.   We note that the ratio $\eta/s$ decreases with increase in temperature in K-matrix formalism in compliance with other model results. While the shear viscosity itself is smaller in case of K-matrix formalism than that of phase shift formalism as shown in Fig.[\ref{shear_pion}(a)] the ratio $\eta/s$ is further reduced by larger entropy density in K-matrix formalism where the additional resonance channel makes extra contribution to the entropy density.  
 
   \begin{figure}[h!]
\vspace{-0.4cm}
\begin{center}
 \includegraphics[width=8cm,height=8cm]{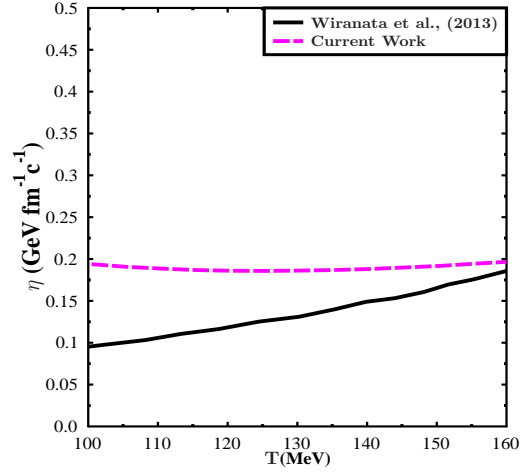}
  \caption{Shear viscosity of the pion gas within thermal averaged relaxation time approximation compared with the Chapman-Enscog approximation estimations of Ref.\cite{Wiranata:2013oaa}. The cross sections are computed using K-matrix parametrization.} 
\label{eta_wira}
  \end{center}
 \end{figure}
 
 Fig.[\ref{eta_wira}] shows the shear viscosity estimate of the pion gas based on relaxation time approximation compared with the Chapman-Enscog approximation\cite{Wiranata:2013oaa}. Both the estimates are based  on K-matrix parametrization of $\pi-\pi$ cross sections. We note that the estimations based on relaxation time approximation are larger than that of Chapman-Enscog approximations. This result is in agreement with the previous results of Ref.\cite{Wiranata:2012br,Wiranata:2012vv}. The difference in the estimate arises due to fact that the Chapman-Enscog approximation features the cross section with the angular weight while the relaxation time approximation lacks this feature. This difference between the estimates is further amplified by the fact that we have used the thermal averaged relaxation times (Eq. (\ref{therm_av})) to estimate the transport coefficients.

  \begin{figure}[h!]
\vspace{-0.4cm}
\begin{center}
\begin{tabular}{c c}
 \includegraphics[width=8cm,height=8cm]{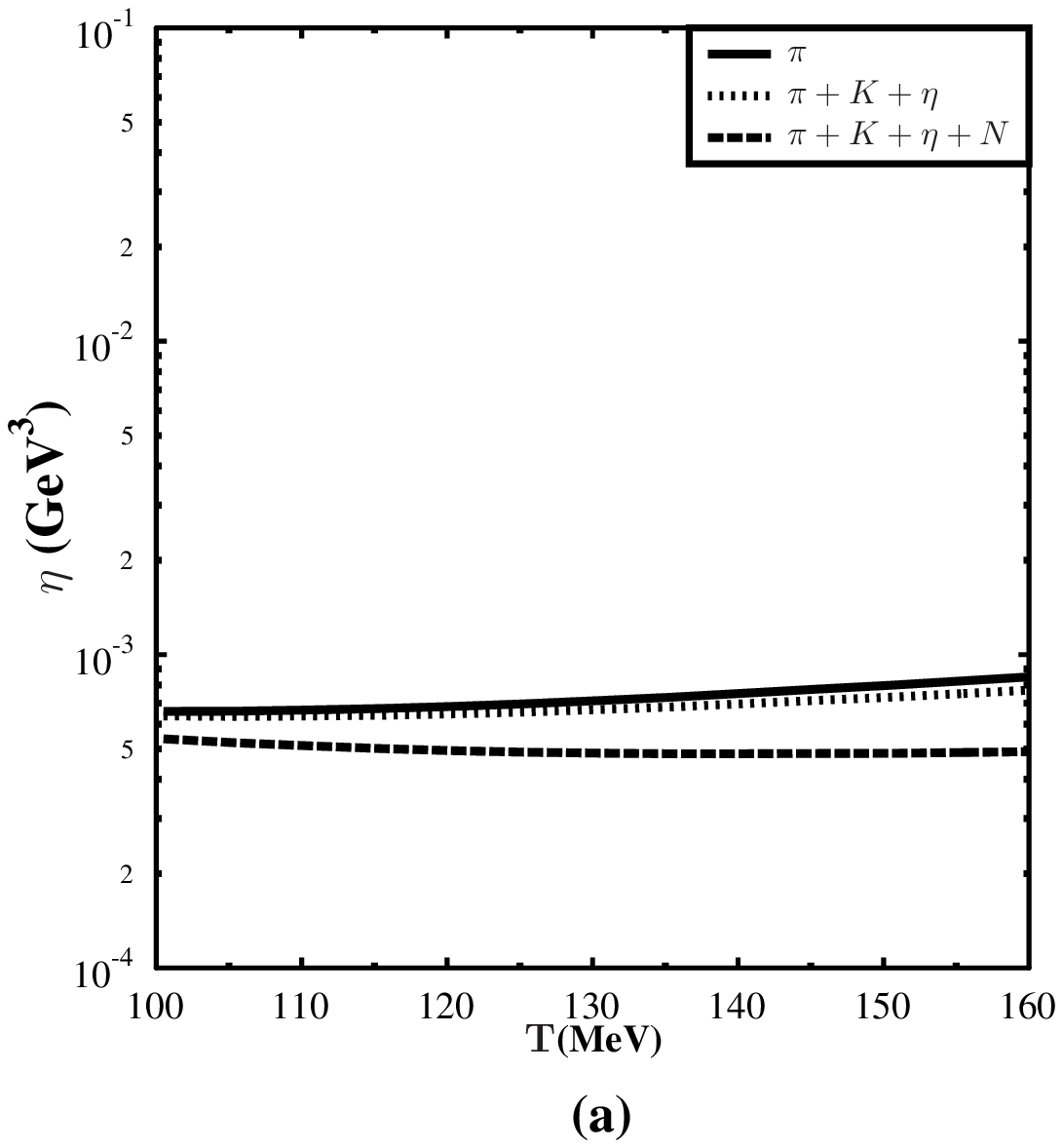}&
  \includegraphics[width=8cm,height=8cm]{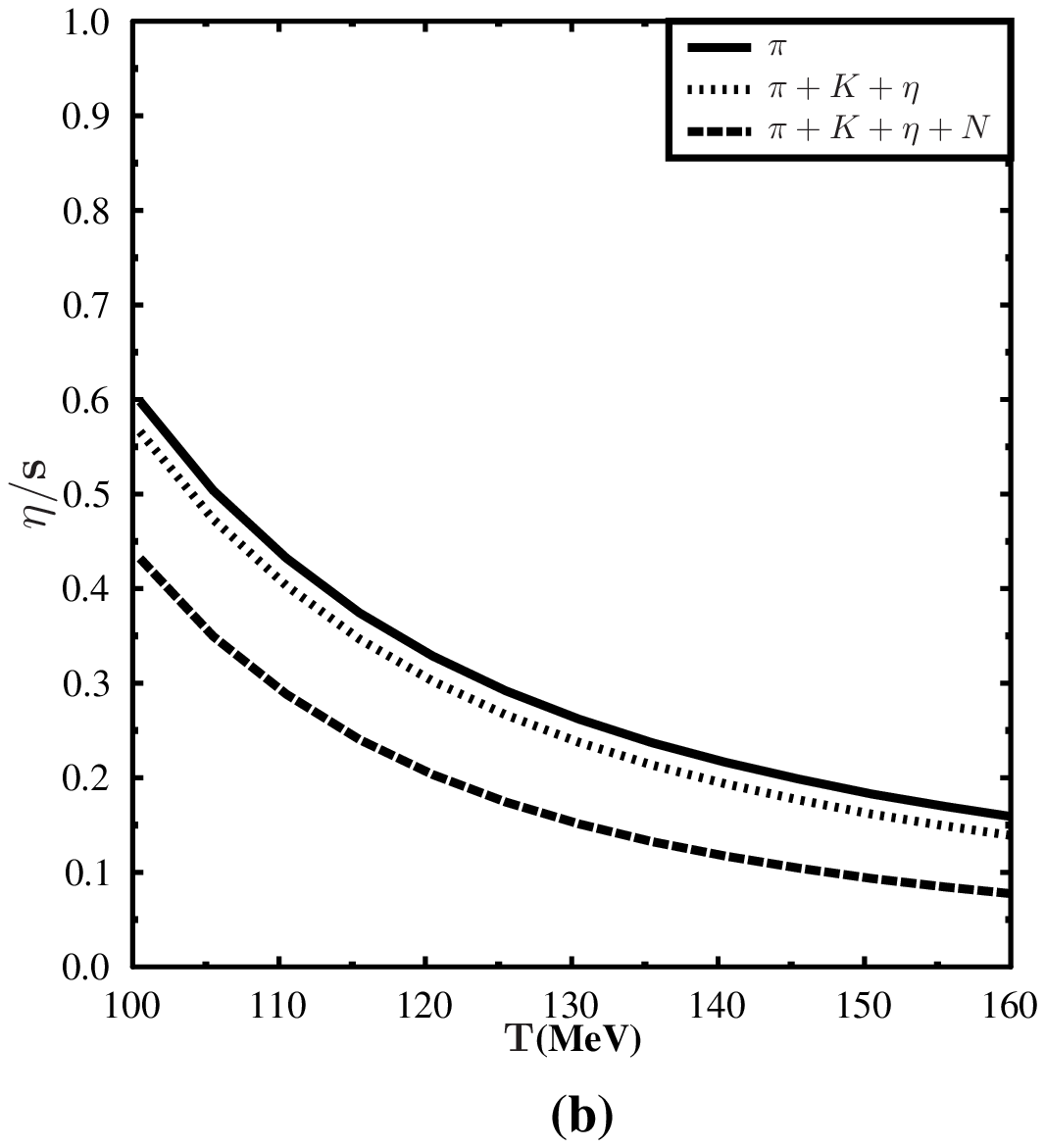}\\
  \end{tabular}
  \caption{Left panel shows shear viscosity estimate for the gas of pions and the mixture of  $\pi-K-\eta$ as well as $\pi-K-\eta-N$ using K-matrix parametrization of cross section. Right panel shows ratio $\eta/s$.  $K-K$, $\eta-\eta$ and $N-N$ interactions are neglected.} 
\label{shear_mix}
  \end{center}
 \end{figure}
 
 Fig.[\ref{shear_mix}(a)] shows the shear viscosity estimate of the gas of pions and the mixture of  $\pi-K-\eta$ as well as $\pi-K-\eta-N$ using K-matrix parametrization of cross section. We note that additional $\pi-K$, $\pi-\eta$ and $\pi-N$ interactions reduces the shear viscosity. Fig.[\ref{shear_mix}(b)] shows the ratio of $\eta/s$ for different hadronic mixtures. This ratio is significantly reduced for $\pi-K-\eta-N$ mixture as compare to the pure pion gas. Thus the heavier hadrons play significant role in the transport phenomena of the hadronic fluid.

 \begin{figure}[h!]
\vspace{-0.4cm}
\begin{center}
\begin{tabular}{c c}
 \includegraphics[width=8cm,height=8cm]{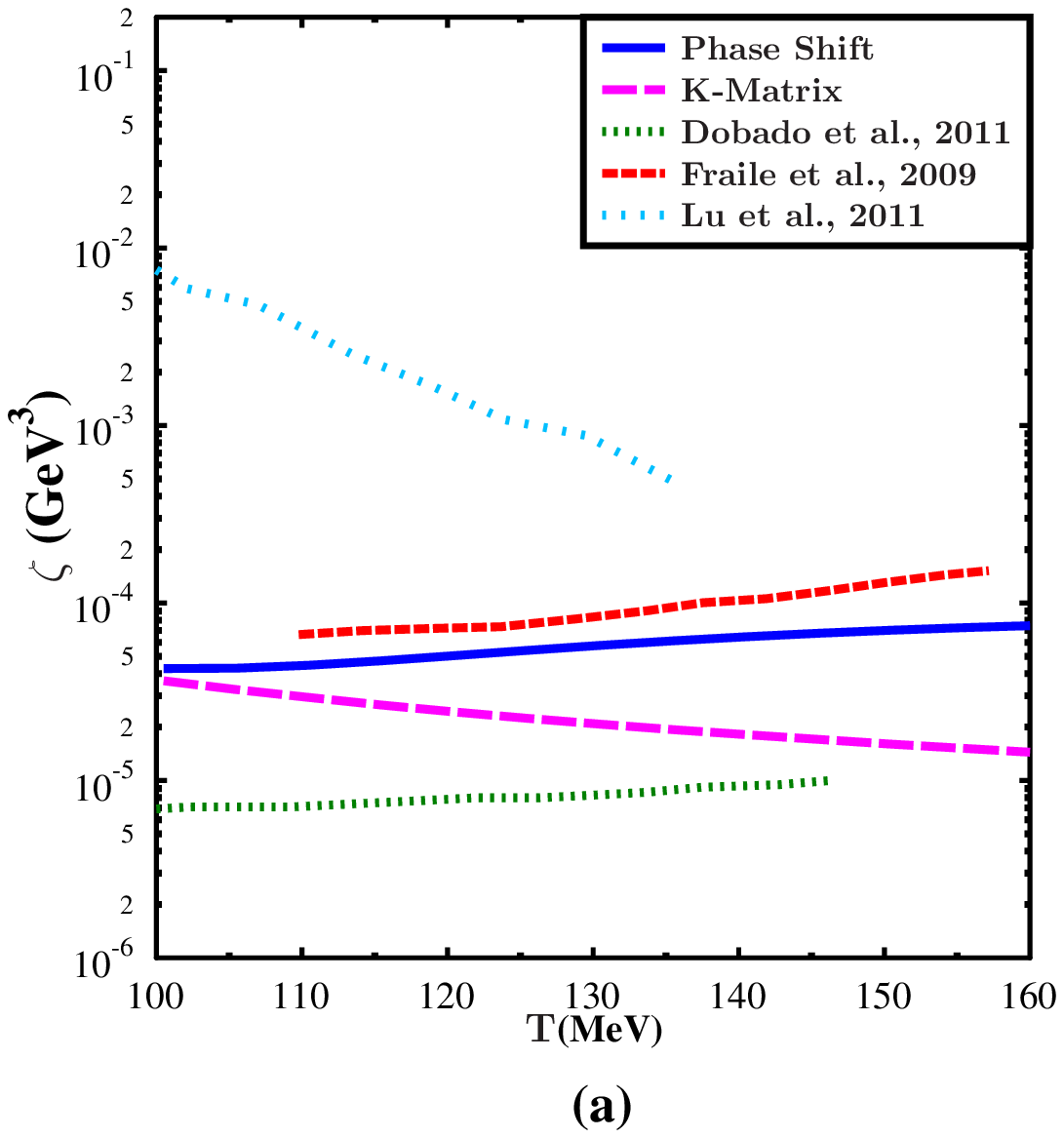}&
  \includegraphics[width=8cm,height=8cm]{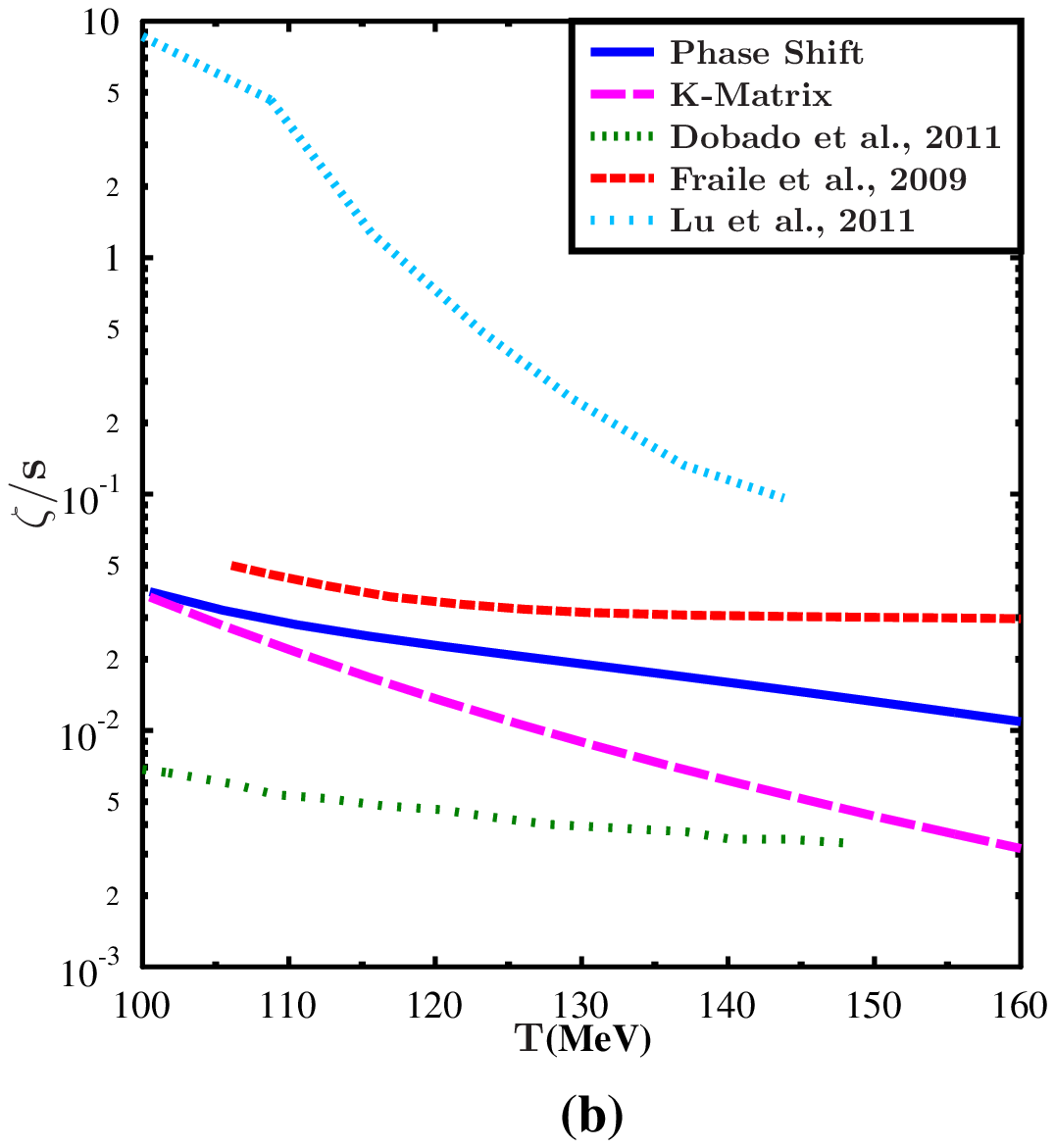}\\
  \end{tabular}
  \caption{Left panel shows bulk viscosity estimate for the gas of pions and the mixture of  $\pi-K-\eta-N$ using K-matrix parametrization of cross sectio. Right panel shows ratio $\eta/s$.} 
\label{zeta_pion}
  \end{center}
 \end{figure}
 
  \begin{figure}[h!]
\vspace{-0.4cm}
\begin{center}
\begin{tabular}{c c}
 \includegraphics[width=8cm,height=8cm]{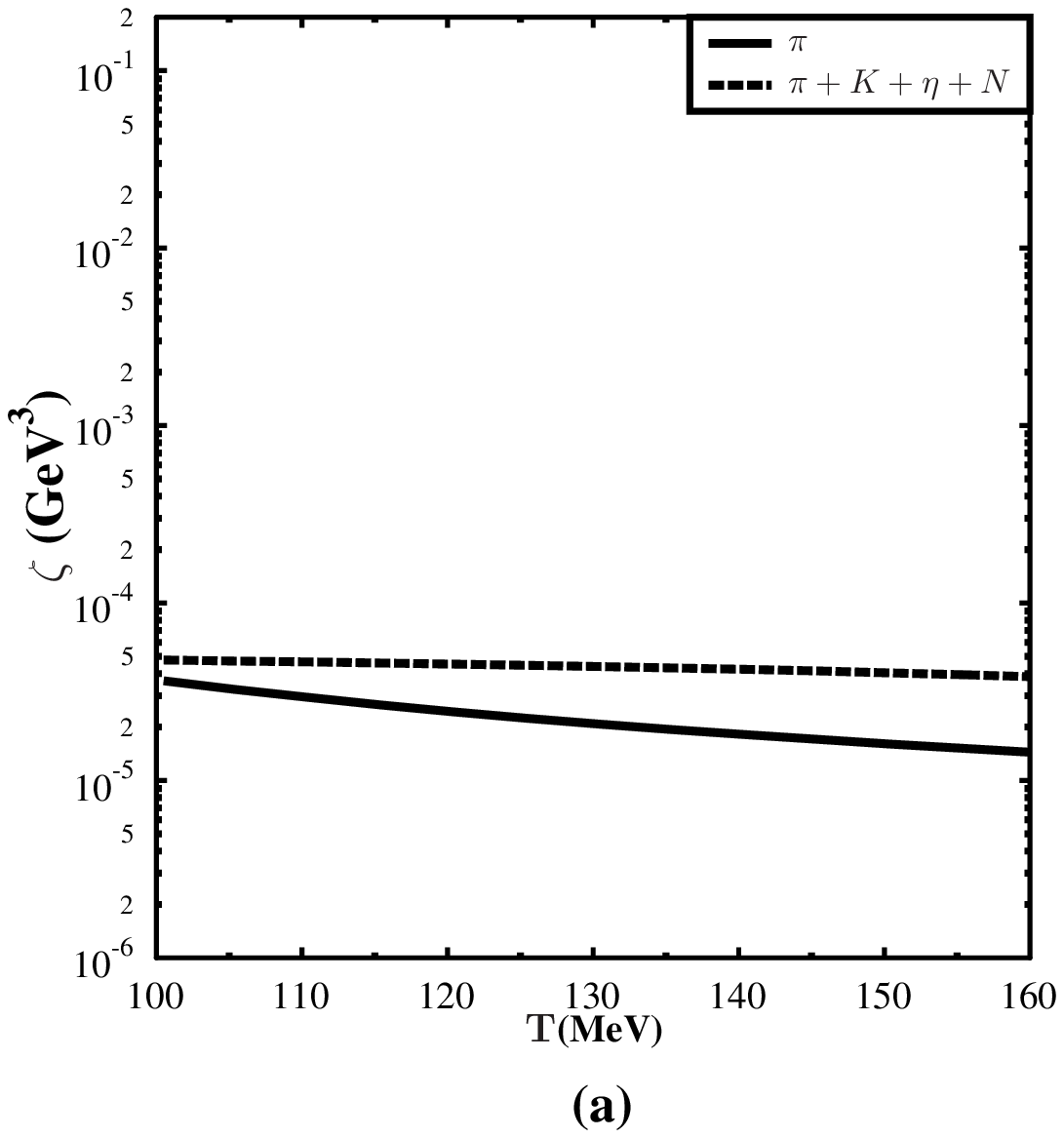}&
  \includegraphics[width=8cm,height=8cm]{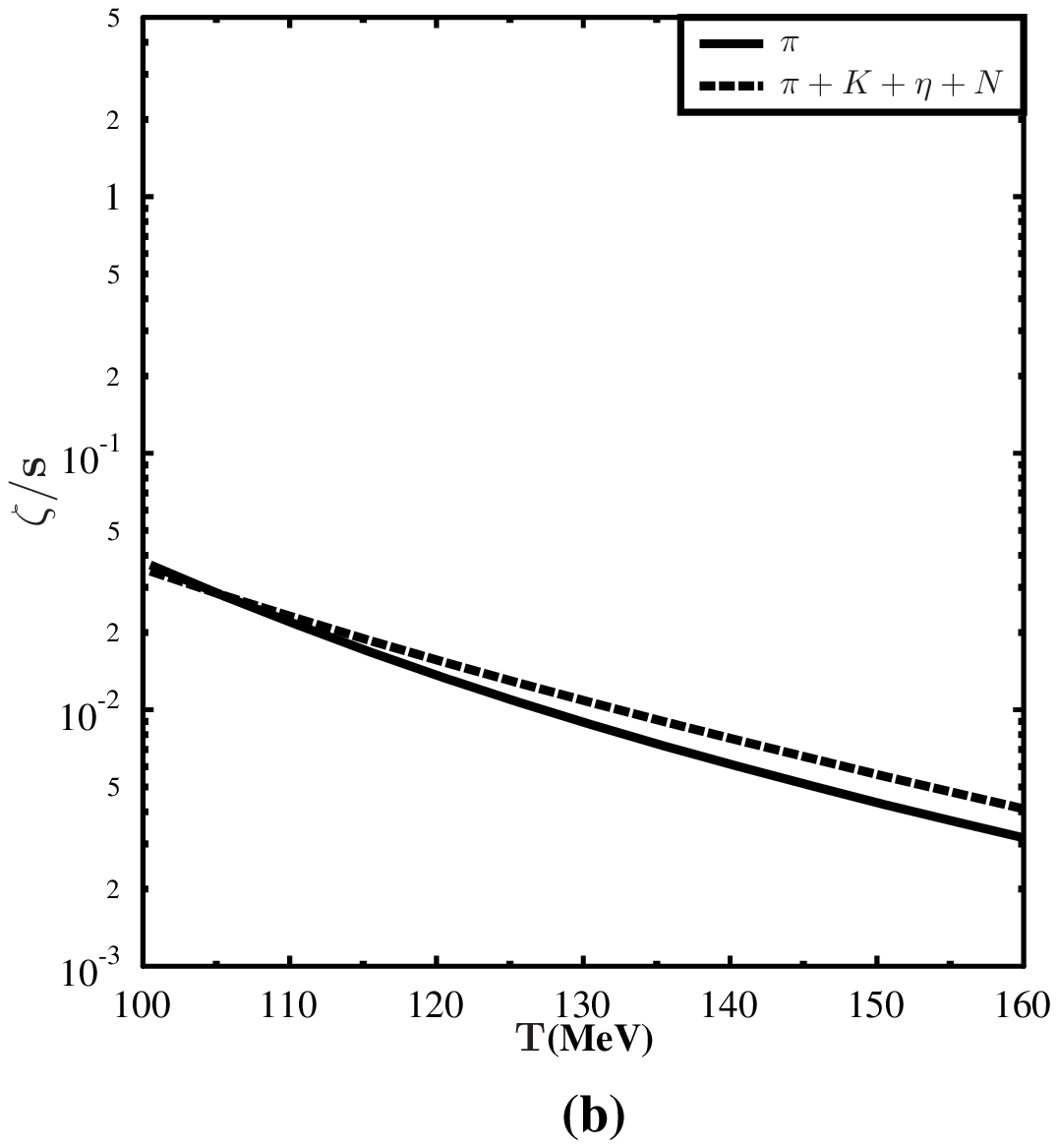}\\
  \end{tabular}
  \caption{Left panel shows bulk viscosity estimate for pion gas obtained using phase shift parametrization used in this work  compared to other estimations\cite{FernandezFraile:2009mi,Dobado:2011qu,Lu:2011df}. Right panel shows ratio $\zeta/s$ for pion gas compared to other works.} 
\label{zeta_mix}
  \end{center}
 \end{figure}
  
 Fig. [\ref{zeta_pion}(a)] shows bulk viscosity  as a function of temperature estimated using phase shift (solid blue) and K-matrix (dashed magenta) cross section parametrization. We note that bulk viscosity  estimated using K-matrix formalism is of the same order of magnitude as that of ChPT results which take into account only elastic $\pi-\pi$ scattering\cite{FernandezFraile:2009mi,Dobado:2011qu}. The bulk viscosity estimated taking into account inelastic process\cite{Lu:2011df} is two order of magnitude higher than K-matrix estimation with elastic scattering processes. Thus inelastic  processes contribute significantly to the bulk viscosity. Fig.[\ref{zeta_pion}(b)] shows the ratio $\zeta/s$ estimated in our model and compared with other model calculations. The ratio $\zeta/s$ decreases with increase in temperature. The higher value of the ratio at low temperature is due to strong conformal symmetry breaking at low temperature ($T\sim m_{\pi}$) which depends on $m_\pi/T$.  We further note that the ratio is significantly smaller when only elastic processes are considered as compared to the case when inelastic processes are also taken into account\cite{Lu:2011df}. The effect of additional interactions of pions with kaons, $\eta$ mesons and nucleons can be seen in Fig.[\ref{zeta_mix}]. We note that the bulk viscosity is larger for $\pi-K-\eta-N$ mixture as compared to pure pion gas. This is due to additional conformal symmetry breaking measure ($\epsilon-3P$) which is higher for the $\pi-K-\eta-N$ mixture as compared to pure pion gas.

 \begin{figure}[h!]
\vspace{-0.4cm}
\begin{center}
\begin{tabular}{c c}
 \includegraphics[width=8cm,height=8cm]{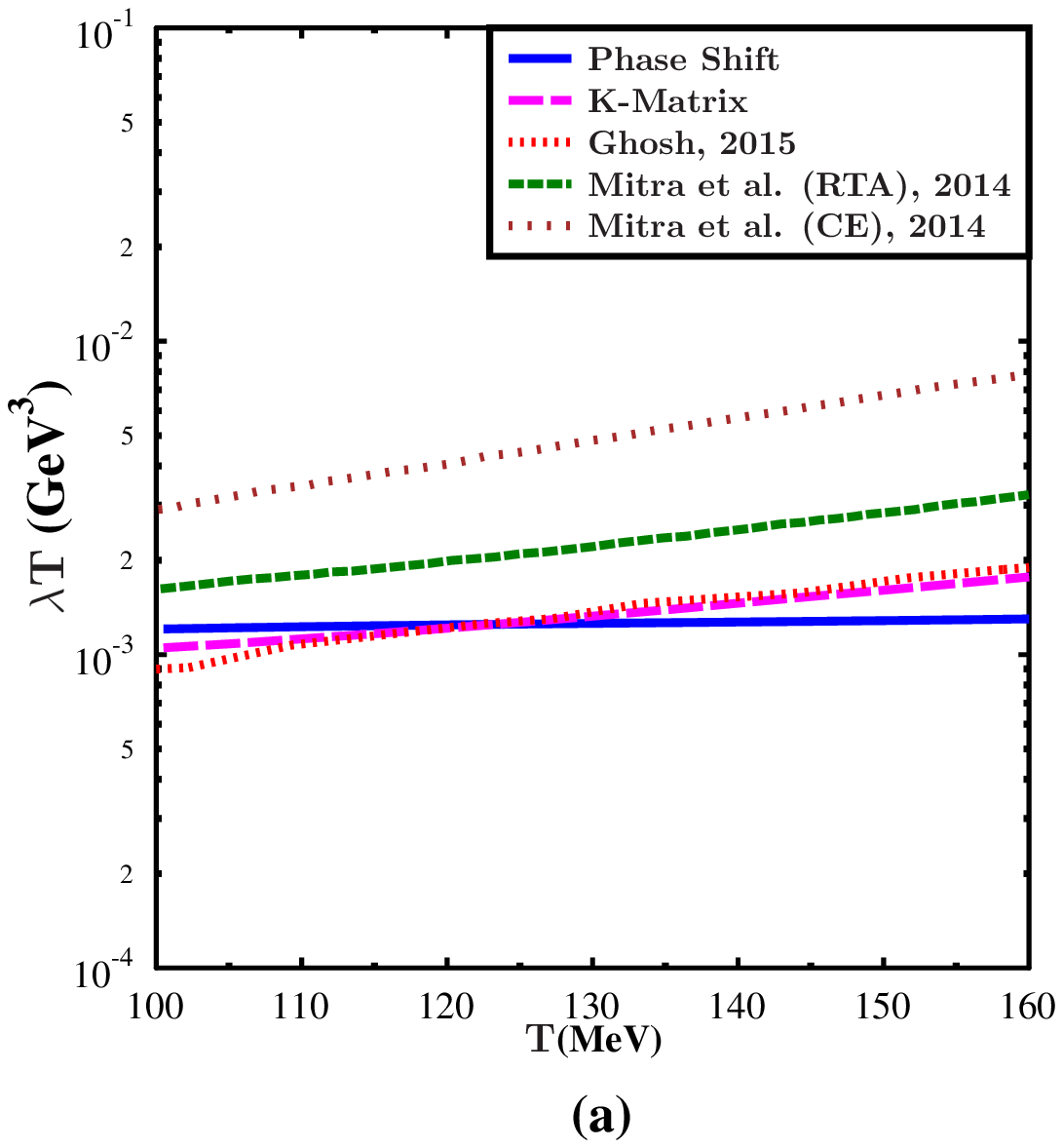}&
  \includegraphics[width=8cm,height=8cm]{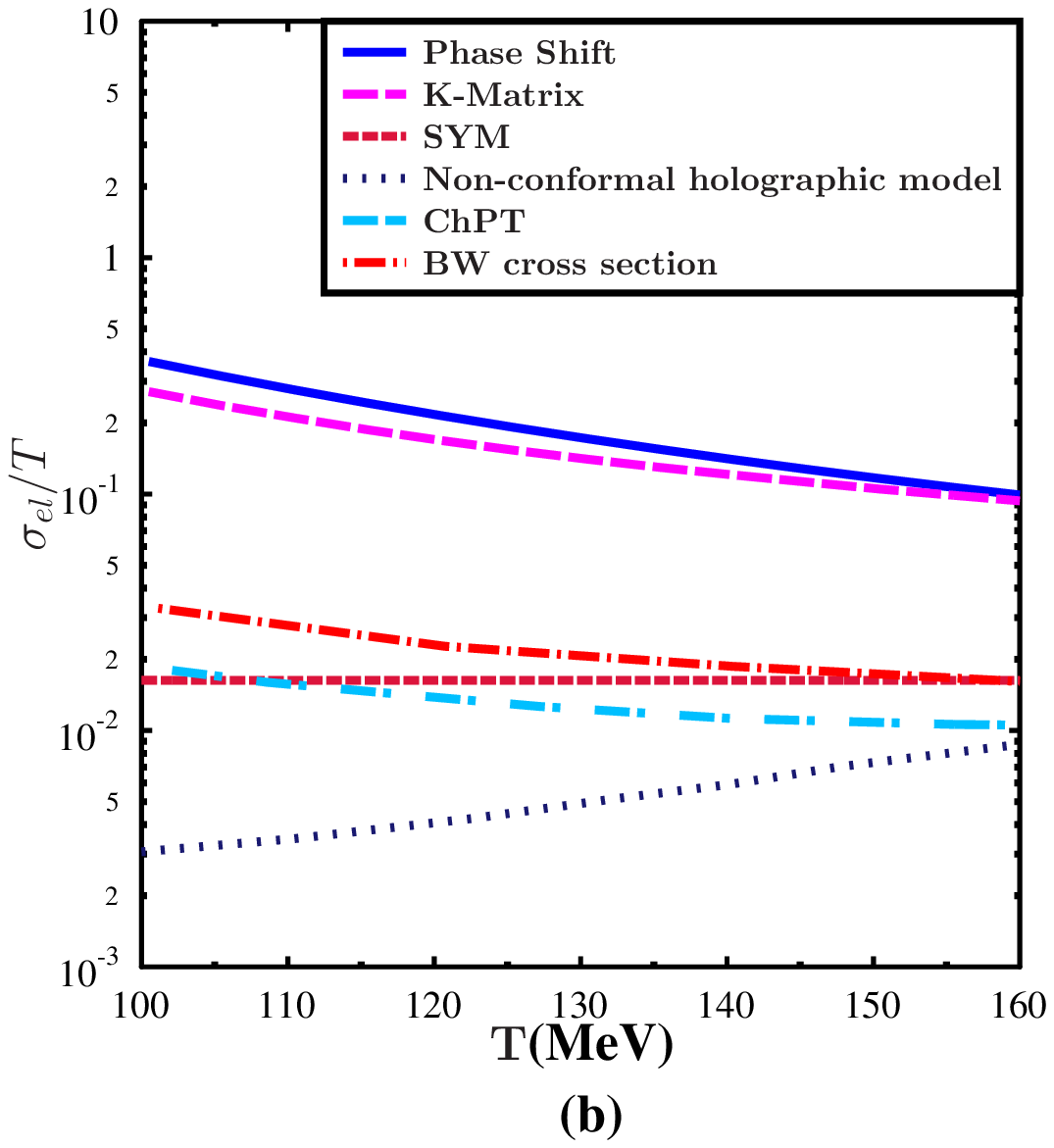}\\
  \end{tabular}
  \caption{Left panel shows thermal conductivity estimate for pion gas obtained using phase shift parametrization used in this work  compared to other estimations\cite{Ghosh:2015mba,Mitra:2014dia}. Right panel shows electrical conductivity for pion gas compared to other works\cite{Greif:2016skc,CaronHuot:2006te,FernandezFraile:2005ka,Finazzo:2013efa}. The sky blue curve correspond to ChPT-based analysis uses the Green-Kubo formalism which expresses conductivity in terms of spectral function\cite{FernandezFraile:2005ka}. The conductivity is extracted from the spectral function by identifying the dominant diagrams in a low energy and low temperature expansion.  Brown dashed curve correspond to conformal Yang-Mills plasma\cite{CaronHuot:2006te} while midnight blue dashed curve correspond to non-conformal holographic model\cite{Finazzo:2013efa}.} 
\label{cond_pion}
  \end{center}
 \end{figure}
 
 \begin{figure}[h!]
\vspace{-0.4cm}
\begin{center}
\begin{tabular}{c c}
 \includegraphics[width=8cm,height=8cm]{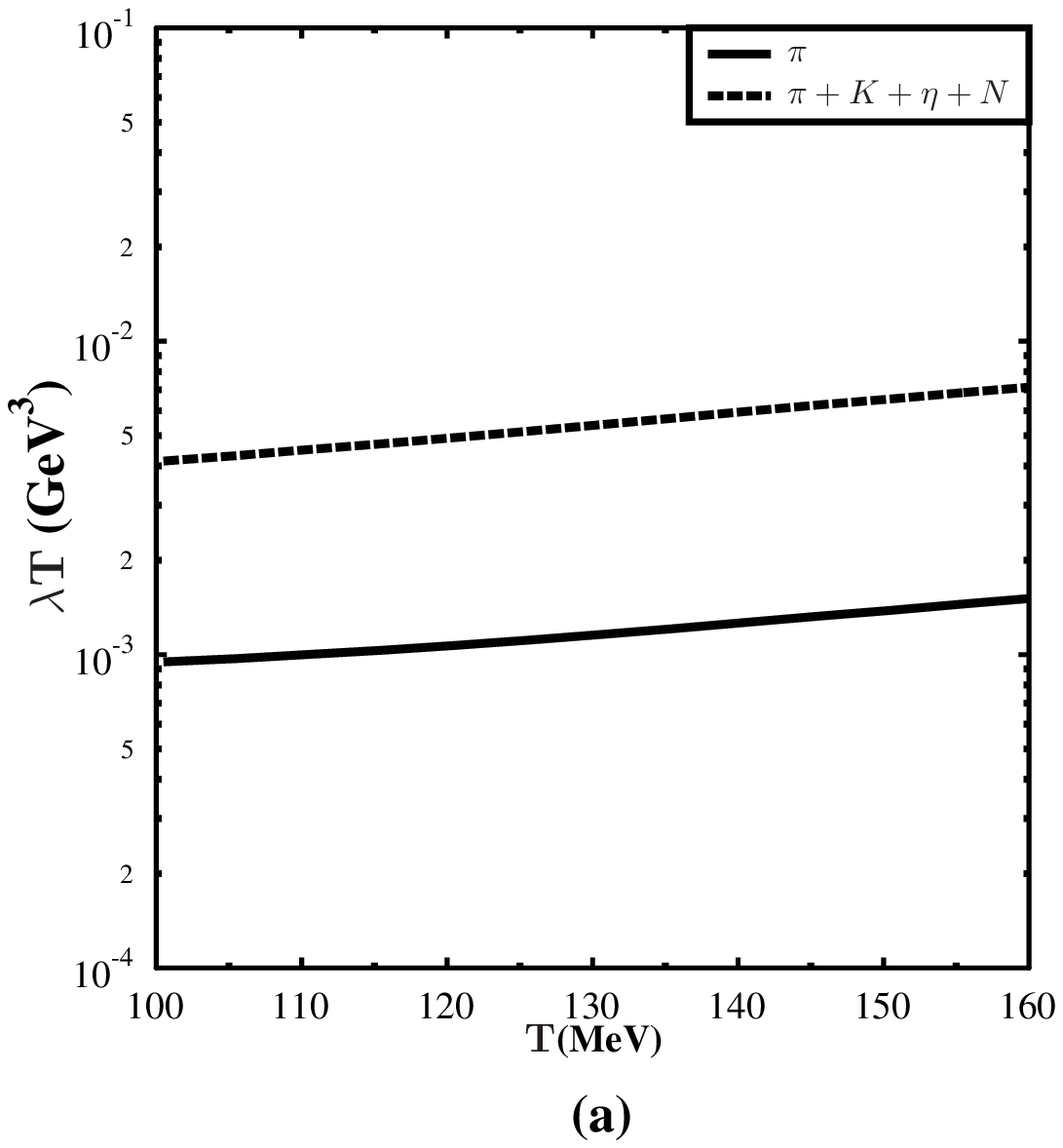}&
  \includegraphics[width=8cm,height=8cm]{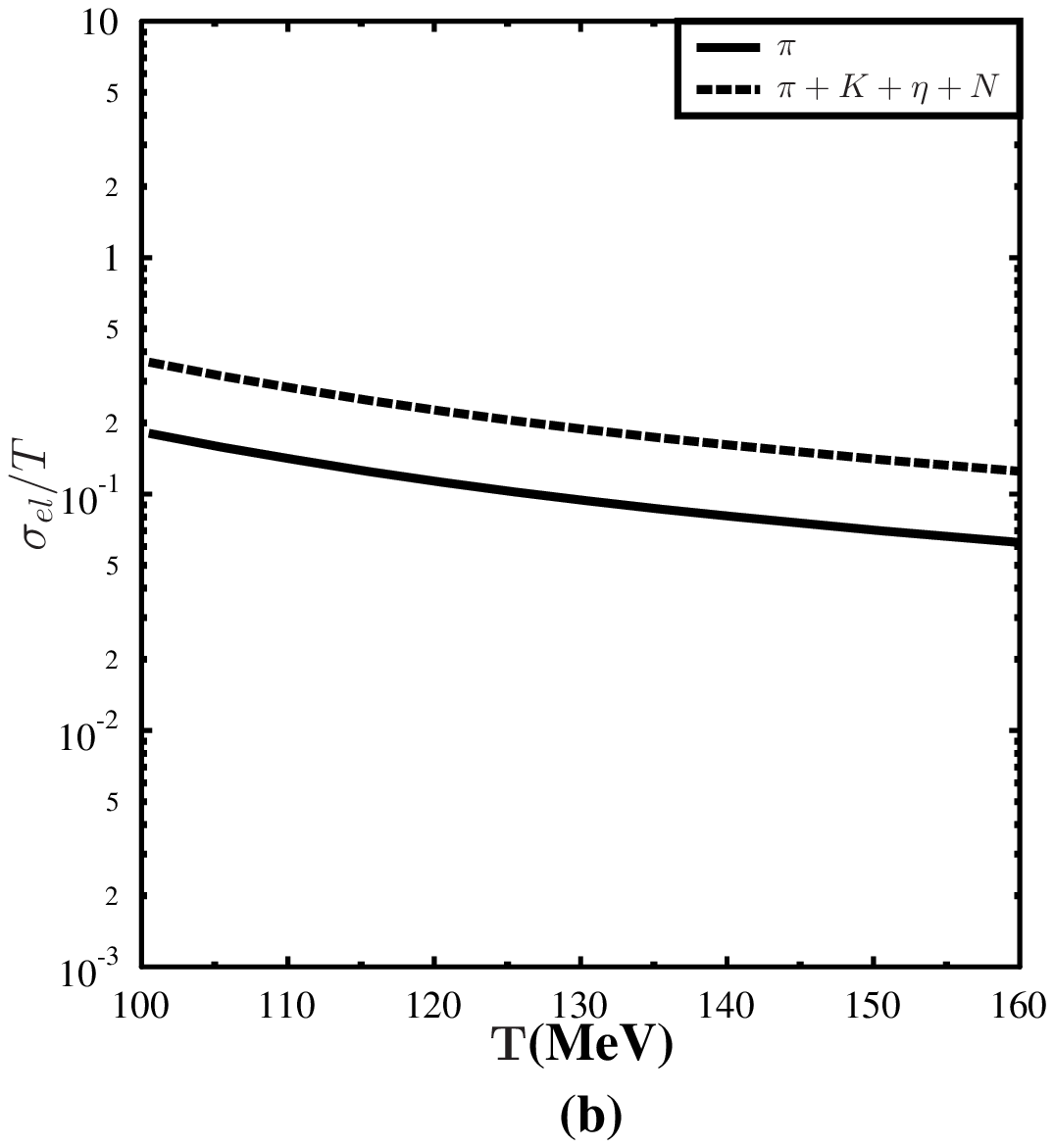}\\
  \end{tabular}
  \caption{Left panel shows thermal conductivity  for the gas of pions  and for the mixture $\pi-K-\eta-N$. Right panel shows electrical conductivity for the gas of pions as well as $\pi-K-\eta-N$. The cross sections has been estimated using K-matrix parametrization scheme.} 
\label{cond_mix}
  \end{center}
 \end{figure}
 
Fig. [\ref{cond_pion}(a)] shows thermal conductivity (times temperature) estimated using phase shift (solid blue) and K-matrix (dashed magenta) cross section parametrization. We have compared our results with that of Ref.\cite{Ghosh:2015mba,Mitra:2014dia}. The red dashed curve correspond to the $\pi-\pi$ inelastic scattering process of Ref.\cite{Ghosh:2015mba}. The green dashed curve corresponds to thermal conductivity estimations based on  the relaxation time approximation while the brown dashed curve correspond to Chapman-Enscog approximation\cite{Mitra:2014dia}. The cross sections are obtained from elastic $\pi-\pi$ scattering  involving $\rho$ and $\sigma$ meson exchange. We note that $\lambda T$ is of same order of magnitude  in K-matrix formalism as compared to that of Ref.\cite{Ghosh:2015mba} but smaller as compared to that of Ref.\cite{Mitra:2014dia}. It is to be noted that the green curve and magenta curve (our work) corresponds to averaged thermal relaxation times. The difference in the estimations can be attributed to the fact that the Ref.\cite{Mitra:2014dia} takes into account only $\sigma$ and $\rho$ meson contribution in $\pi-\pi$ scattering. Further they incorporate the medium effects through thermal one-loop self energies. Fig.[\ref{cond_mix}(a)] shows effect of heavier hadronic interactions with pions on the thermal conductivity. The thermal conductivity is larger for $\pi-K-\eta-N$ mixture as compared to pure pion gas. This rise is due to higher value of the heat function $h$ in case of hadronic mixture as compare to that of pure pion gas.
 
Fig. [\ref{cond_pion}(b)] shows electrical conductivity (times temperature) estimated using phase shift (solid blue) and K-matrix (dashed magenta) cross section parametrization. We note that K-matrix formalism gives relatively large conductivity as compared to other model results\cite{Greif:2016skc,CaronHuot:2006te,FernandezFraile:2005ka,Finazzo:2013efa}. The large conductivity is our model can be attributed to relatively smaller cross section as compared to other models. For instance,  the red curve corresponds to pions $\pi^{\pm},\pi^0$ interacting via isotropic resonance cross section computed using Breit-Wigner form of scattering amplitude for the dominant $\rho$ meson decay channel\cite{Greif:2016skc}. In this case the $\sigma_{tot}\sim30$ mb whereas in our model $\sigma_{tot}\sim 10$ mb. We further note that $\sigma/T$ estimated in our model approaches similar minimum as that of \cite{Greif:2016skc} which is expected at hadron-QGP phase transition. Fig.[\ref{cond_mix}(b)] shows the effect of  heavier hadronic interactions with pions on the electrical conductivity. The electrical conductivity is larger for $\pi-K-\eta-N$ mixture as compared to pure pion gas. This rise may be attributed to the fact that more charged particles are available in the mixture of the hadron gas which contribute to the electrical conductivity.

%\newpage
 \section{summary}
 In this work we estimated the transport coefficients, $viz.$, shear and bulk viscosities as well as thermal and electrical conductivities, of  pion gas using K-matrix formalism to compute the $\pi\pi$ cross sections. This formalism  incorporate multiple heavy resonance channels to compute the scattering amplitudes while simultaneously preserving the unitarity of S-matrix. Our estimations are in reasonable agreement with other results based on various model calculations. The shear viscosity coefficient is smaller compared to the results of chiral perturbation theory\cite{Lang:2012tt,FernandezFraile:2009mi} while it is larger when compared with Chapman-Enscog method of Ref.\cite{Wiranata:2013oaa}. The bulk viscosity  estimated using K-matrix formalism is of the same order of magnitude as that of ChPT results which take into account only elastic $\pi-\pi$ scattering\cite{FernandezFraile:2009mi,Dobado:2011qu} while it is two order of magnitude smaller when inelastic processes are taken in to account\cite{Lu:2011df}. Thus inelastic processes might play important role in determining the bulk viscosity. The thermal conductivity coefficient estimations based on K-matrix cross section are also found to be in reasonable agreement with previous results\cite{Ghosh:2015mba,Mitra:2014dia}. The electrical conductivity coefficient estimated using K-matrix formalism gives relatively large value as compared to other model results\cite{Greif:2016skc,CaronHuot:2006te,FernandezFraile:2005ka,Finazzo:2013efa}. The large conductivity in our model can be attributed to relatively smaller cross section as compared to other models.  We further studied the effect of higher order corrections, $viz.$ $\pi-K$, $\pi-\eta$ and $\pi-N$ interactions, on transport coefficients. We found that additional interactions reduces the shear viscosity coefficient but increases bulk viscosity coefficient as well as thermal and electrical conductivities. It is to be noted that our calculations were restricted to elastic processes. But inelastic processes are important as far as the bulk viscosity coefficient is concerned.  Further, our formalism can be reasonably extended to include repulsive interactions as well as other higher order effects like  baryon interactions which are of phenomenological importance in the context of off-central heavy ion collision experiments which are capable of producing baryon rich fluid. Finally, it would be very interesting to see the effect of 3-pion interactions\cite{Doring:2006ue,Mai:2018djl} on the transport coefficients.

 \section{Appendix}
 In this section we just recapitulate the phase shift and K-matrix parametrization given in Ref.\cite{Wiranata:2013oaa} to make the paper readable and self contained.
 
 \subsection{Phase shift parametrization}
 The differential scattering cross section for a process $ab\longrightarrow cd$ can be written in terms of scattering amplitude $f(\sqrt{s},\theta)$ as
 \be
 \frac{d\sigma}{d\Omega}_{CM}=\mid f(\sqrt{s},\theta)\mid^{2}
 \ee
 The scattering matrix $S$ is defined as an overlap matrix between initial and final states for the process $ab\longrightarrow cd$. It can be conveniently written as
 \be
 S=I+2ikf=I+2iT
 \ee
 
 The scattering amplitude $f$ can be expressed in terms of interaction matrix $T$  as
 \be
 f(\sqrt{s},\theta)=\frac{1}{q_{CM}}\sum_{l}(2l+1)T^{l}(s)P_{l}(cos\theta)
 \ee
 Thus the scattering cross section becomes
 \be
 \frac{d\sigma}{d\Omega}_{CM}=\frac{2}{q_{CM}^{2}}\sum_{l}(2l+1)\mid T^{l}(s)\mid^{2}
 \label{diffsigma}
 \ee
 where we have used the orthonormality of Legendre polynomials. In terms of phase shift $\d_{l}$ the interaction matrix can be written as
 \be
 T^{l}=e^{i\d_{l}}\text{sin}(\d_{l})
 \ee
 The total scattering cross section for a given channel is

\be
\sigma^{I}_{l}(\sqrt{s})=\frac{8\pi}{q^{2}}(2l+1)\text{sin}^{2}\d^{I}_{l}(\sqrt{s})
\ee

For the case of elastic $\pi\pi\longrightarrow\pi\pi$ scattering we consider the three channels, $\d^{0}_{0},\d^{2}_{0}$ and $\d^{1}_{1}$ in which the phase shifts are known experimentally. The parametrization for these channels are respectively given by\cite{Bertsch:1987ux}
\be
\d^{0}_{0}=\frac{\pi}{2}+\text{arctan}\bigg(\frac{\sqrt{s}-m_{\sigma}}{\Gamma_{\sigma}/2}\bigg);\: \Gamma_{\sigma}=2.06q_{CM};\: m_{\sigma}=5.8m_{\pi}
\ee

\be
\d^{2}_{0}=-0.12\frac{q_{\text{CM}}}{m_{\pi}}
\ee

\be
\d^{1}_{1}=\frac{\pi}{2}+\text{arctan}\bigg(\frac{\sqrt{s}-m_{\rho}}{\Gamma_{\rho}/2}\bigg);\: \Gamma_{\rho}=0.095q_{\text{CM}}\bigg[\frac{q_{\text{CM}}/m_{\pi}}{(1+(q_{\text{CM}}/m_{\rho})^2)}\bigg]^2;\: m_{\rho}=5.53m_{\pi}
\ee

The isospin averaged cross section over these three channels is given by
\be
\sigma_{av}^{\pi\pi}=\frac{1}{9}\sigma^{0}+\frac{5}{9}\sigma^{2}+\frac{1}{3}\sigma^{1}
\ee
 
Thus the isospin averaged $\pi\pi$ total cross section is

\be
\sigma_{av}^{\pi\pi}(\sqrt{s})=\frac{8\pi}{q^{2}_{\text{CM}}}\bigg(\frac{1}{9}\text{sin}^{2}\d^{0}_{0}+\frac{5}{9}\text{sin}^{2}\d^{2}_{0}+\frac{1}{3}3\:\text{sin}^{2}\d^{1}_{1}\bigg)
\ee

 \subsection{K-Matrix parametrization}
 The unitarity of $S$ matrix can be expressed by relation
 \be
 SS^{\dag}=S^{\dag}S=I
 \ee
 
Using this unitarity relation one can define hermitian matrix $K^{-1}$ as
\be
K^{-1}=T^{-1}+iI,\: K^\dag=K
\ee
 
 One can then express $T$ matrix in terms of $K$ matrix through relations
 \be
 \text{Re} T=K(I+K^2)^{-1};\:\text{Im} T=K^2(I+K^2)^{-1}
 \label{ktrel}
 \ee
 The resonances in K-matrix formalism appear as a sum of poles
 \be
 K_{ab\longrightarrow cd}=\sum_{R}\frac{g_{R\longrightarrow ab}g_{R\longrightarrow cd}}{m_{R}^{2}-s}
 \ee where $g^2=m_{R}\Gamma_{R}$. The partial decay widths are given by (for the decay $R\longrightarrow ab$)
 \be
 \Gamma_{R\longrightarrow ab}(\sqrt{s})=\Gamma^{0}_{R\longrightarrow ab}\frac{q_{CM}}{q_{CM}^{'}}\frac{m_{R}}{\sqrt{s}}[B^{l}(q_{CM},q_{CM}^{'}]^2
 \ee
 where $q_{CM}^{'}=\frac{1}{2}\sqrt{m_{R}^2-4m_{a}m_{b}}$ is the break up momentum and $B^{l}$'s are Blatt-Weisskopf barrier factors\cite{VonHippel:1972fg}.  Thus once we know $K$-matrix one can obtain $T$-matrix using relations (\ref{ktrel}) and hence the cross sections using (\ref{diffsigma}).
 
If  $\pi-\pi$ scattering occurs through two resonance channels  with mass $m_1$ and $m_2$ then the K-matrix can be written as

\begin{equation}
 K=\frac{m_1\Gamma_1(\sqrt{s})}{m_1^2-s}+\frac{m_2\Gamma_2(\sqrt{s})}{m_2^2-s},
\end{equation}

 Corresponding T-matrix is\cite{Dash:2018mep}
\begin{equation}
  T = \frac{m_1\Gamma_{1}(\sqrt{s})}{(m_1^2-s)-im_1\Gamma_1(\sqrt{s})-i\frac{m_1^2-s}{m_2^2-s}m_2\Gamma_2(\sqrt{s})}+\\
  \frac{m_2\Gamma_{2}(\sqrt{s})}{(m_2^2-s)-im_2\Gamma_2(\sqrt{s})-i\frac{m_2^2-s}{m_1^2-s}m_1\Gamma_1(\sqrt{s})}.
\end{equation}

If these two resonances are separated  then $K$-matrix is dominated either by $m_1$ or $m_2$ depending on whether $\sqrt{s}$ is
near $m_1$ or $m_2$. The transition amplitude can be written as
\begin{equation}
 T \approx \frac{m_1\Gamma_{1}(\sqrt{s})}{(m_1^2-s)-im_1\Gamma_1(\sqrt{s})}+\frac{m_2\Gamma_{2}(\sqrt{s})}{(m_2^2-s)-im_2\Gamma_2(\sqrt{s})},
\end{equation}

\begin{figure}[h!]
\vspace{-0.4cm}
\begin{center}
\begin{tabular}{c c}
 \includegraphics[width=8cm,height=8cm]{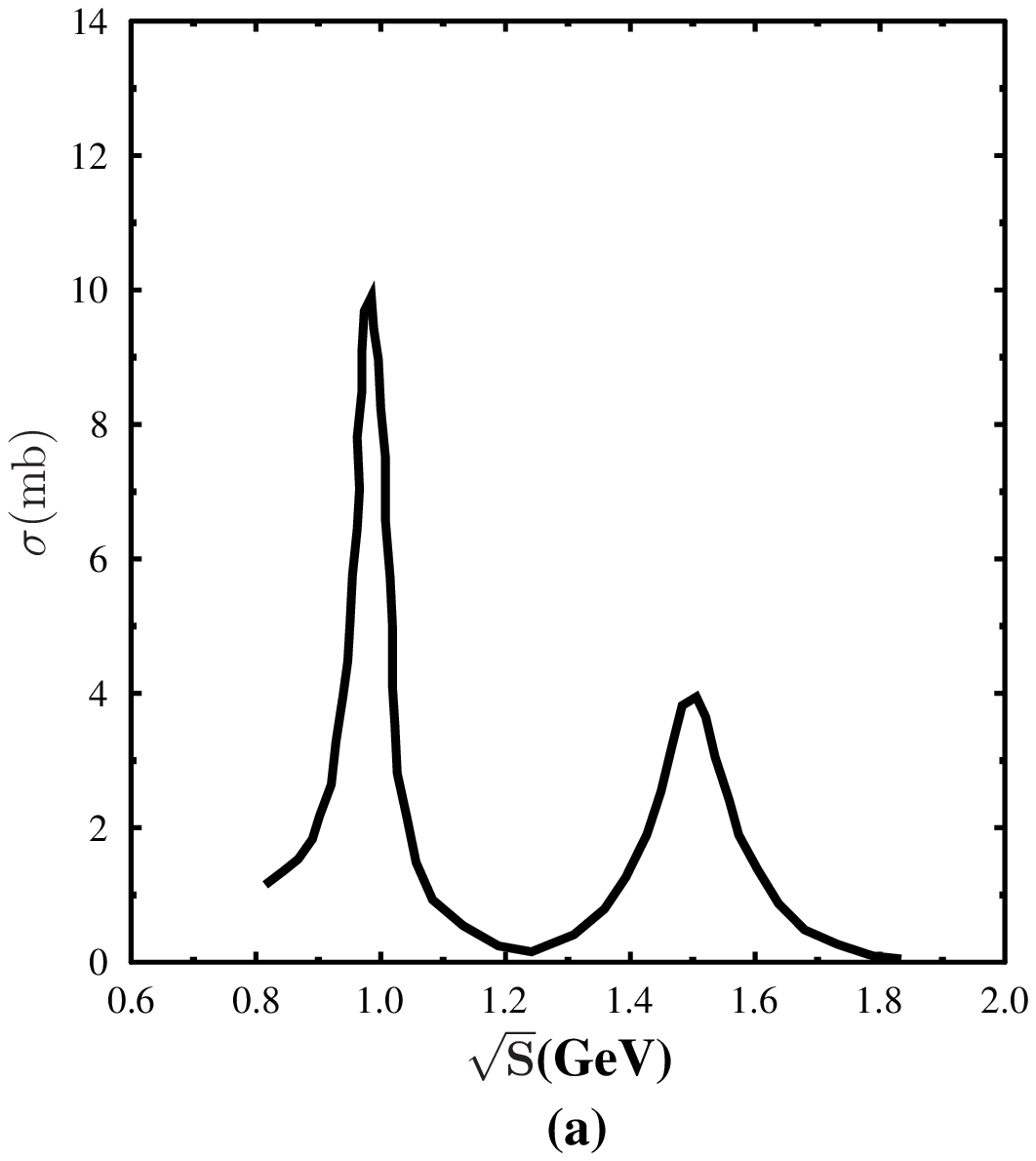}&
  \includegraphics[width=8cm,height=8cm]{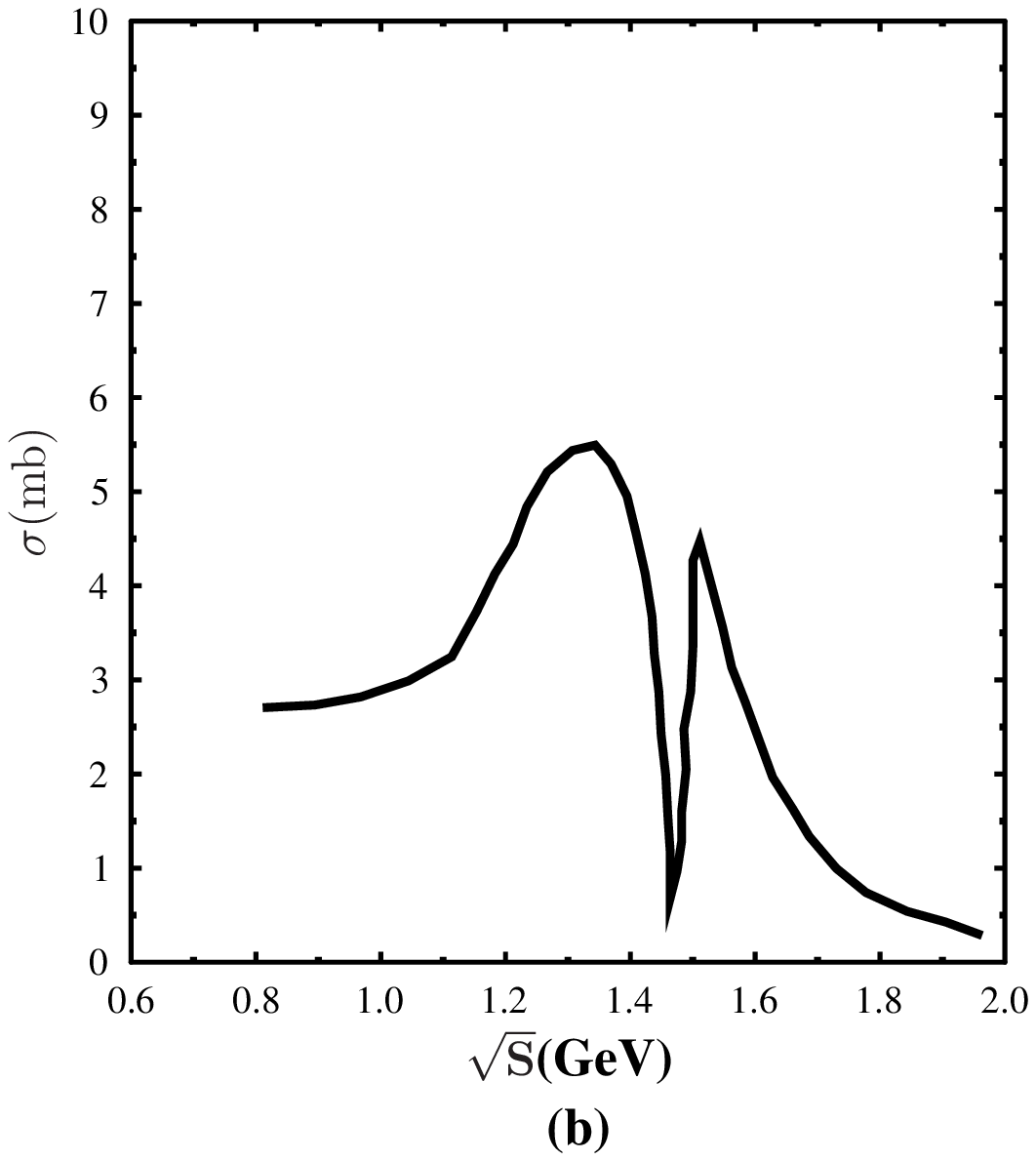}\\
  \end{tabular}
  \caption{Left panel shows the total cross section computed using K-matrix parametrization  for two separated resonances $f_0(980)$ and $f_{2}^{'}(1525)$. Right panel shows the cross section computed using k-matrix parametrization for two overlapping resonances $f_{02}(1370)$  and $f_2^{'}(1525)$.} 
\label{resonance}
  \end{center}
 \end{figure}

 Fig. [\ref{resonance}(a)] shows the total cross section computed using K-matrix parametrization for two separated resonances $f_0(980)$ and $f_{2}^{'}(1525)$.  Fig. [\ref{resonance}(b)] shows the total cross section for two overlapping resonances $f_{02}(1370)$  and $f_2^{'}(1525)$. 
 
 \subsection{List of Resonances}
 
 \begin{table}[h!]
\centering
\begin{tabular}{ |c|c|c|} 
 \hline
 Particle & Mass & Width \\ 
 (MeV) & (MeV) & (MeV) \\ 
  \hline
 $\rho$ & 774 & 150  \\ 
 $\omega$ & 782 & 8  \\ 
 $f_0$ & 980 & 55  \\ 
 $f_2$ & 1270 & 185 \\ 
 $f_{02}$ & 1370 & 200 \\ 
 $\rho_2$ & 1465 & 310  \\ 
 $f'_{2}$ & 1525 & 115  \\ 
 $\rho_3$ & 1690 & 235 \\ 
 $K^*$ & 893 & 50 \\ 
 $ K^*_0$ &1429  &287  \\ 
 $K_2$ &1430  &100  \\
 $K^*_2 $ &1410  &227  \\ 
 $K^*_3 $ &1680  & 323 \\ 
 $a_0 $ &984  &185  \\
 $a_2 $ &1320  &107  \\ 
 $\Delta_{1232} $ &1232  &115  \\
 $ \Delta_{1600}$ &1600  &200  \\
 $\Delta_{1620} $ &1620  &180  \\
 $\Delta_{1700} $ &1700  &300  \\
 $\Delta_{1900} $ &1900  &240  \\
 $\Delta_{1905} $ &1905  &280  \\ 
 $\Delta_{1910} $ &1910  &280  \\
 $\Delta_{1920} $ &1920  &260  \\
 $\Delta_{1930} $ &1930  &360  \\
 $\Delta_{1950} $ &1950  &285  \\
 $N^*_{1440} $ &1440  &200  \\
 $N^*_{1520} $ & 1520 & 125 \\ 
 $N^*_{1535} $ &1535  &150  \\
 $N^*_{1650} $ &1650  &150  \\
 $N^*_{1675} $ &1675  &140  \\
 $N^*_{1680} $ & 1680 &120  \\
 $N^*_{1700} $ &1700  &100  \\
 $N^*_{1710} $ & 1710 &110  \\
  $N^*_{1720} $ &1720  &150  \\ 
  $N^*_{1900} $ & 1900 & 250 \\ 
  $N^*_{1990} $ &1990  & 550 \\ 
 \hline
\end{tabular}
\caption{List of resonances involved in calculation of the cross section.}
\label{table}
\end{table}

 \newpage
 \section*{Acknowledgement}
 GK is financially supported by  the  DST-INSPIRE  faculty  award  under  Grant  No. DST/INSPIRE/04/2017/002293.

\end{document}